\documentclass[reprint, prb,twocolumn,showkeys,showpacs,superscriptaddress]{revtex4-1}
\usepackage{braket}
\usepackage{amsmath,amssymb,times,color,graphicx,multirow,amsfonts,color}
\usepackage[colorlinks=true,citecolor=blue,linkcolor=blue]{hyperref}
\usepackage{physics,bm,here}
\usepackage{silence}
\usepackage{lipsum}

\begin{document}
\title{Symmetry indicators for topological superconductors}
\author{Seishiro Ono}
\affiliation{Institute for Solid State Physics, University of Tokyo, Kashiwa 277-8581, Japan}

\author{Youichi Yanase}
\affiliation{Department of Physics, Graduate School of Science, Kyoto University, Kyoto 606-8502, Japan}

\author{Haruki Watanabe} 
\email{haruki.watanabe@ap.t.u-tokyo.ac.jp}
\affiliation{Department of Applied Physics, University of Tokyo, Tokyo 113-8656, Japan}

\begin{abstract}
The systematic diagnosis of band topology enabled by the method of ``symmetry indicators" underlies the recent advances in the search for new materials realizing topological crystalline insulators. Such an efficient method has been missing for superconductors because the quasi-particle spectrum in the superconducting phase is not usually available. 
In this work, we establish symmetry indicators for weak-coupling superconductors that detect nontrivial topology based on the representations of the metallic band structure in the normal phase assuming a symmetry property of the gap function. We demonstrate the applications of our formulae using examples of tight-binding models and density-functional-theory band structures of realistic materials.
\end{abstract}

\maketitle

\section{Introduction}
In recent years, topological superconductors (SCs) have been actively investigated because Majorana fermions that emerge at vortex cores and on surfaces of topological SCs are promising building blocks of quantum computers~\cite{KITAEV20032,RevModPhys.80.1083,Lian201810003}. 
Intensive experimental efforts have obtained strong indications for topological superconductivity realized in artificial structures by superconducting proximity effect~\cite{He294,Mourik1003,Nadj-Perge602}. 
Further searches for intrinsic topological SCs in crystalline solids are actively ongoing issues. In addition to the topological superconductivity protected by local symmetries~\cite{1367-2630-12-6-065010,doi:10.1063/1.3149495,PhysRevB.78.195125} the topological crystalline superconductivity~\cite{PhysRevB.88.075142,PhysRevB.88.125129,PhysRevB.96.094526,PhysRevLett.111.056403,PhysRevB.90.165114} and higher-order topological superconductivity may be realized in crystalline systems. In previous works a method suitable for a limited set of candidate materials was extensively used~\cite{PhysRevLett.105.097001,PhysRevB.81.220504,PhysRevB.81.134508, PhysRevB.95.224514, PhysRevLett.111.087002}. 
A systematic theory that coherently applies to an enormous number of possible topological SCs is awaited.

Recently, there have been fundamental advances in the method of symmetry indicators~\cite{Po2017,Watanabeeaat8685,PhysRevX.8.031070,QuantitativeMappings,PhysRevX.8.031069} and in a similar formalism~\cite{Bradlyn:2017aa,PhysRevX.7.041069}, which provide an efficient way to diagnose the topology of band insulators and semimetals based on the representations of valence bands at high-symmetry momenta. This scheme can be understood as a generalization of the Fu-Kane formula~\cite{PhysRevB.76.045302} that computes the $\mathbb{Z}_2$-indices in terms of inversion parities to arbitrary (magnetic) space groups~\cite{PG,ITC,Bradley} and a wider class of topologies including higher-order ones~\cite{PhysRevLett.119.246401,PhysRevLett.119.246402,Schindlereaat0346,1709.01929,PhysRevB.97.205135,PhysRevB.97.205136}.  It formed the basis of recent extensive material searches based on the density functional theory (DFT) calculation by several groups that resulted in the discovery of an enormous number of new topological materials~\cite{Tang:2019aa,Vergniory:2019aa,Zhang:2019aa}.  

Up to this moment, however, symmetry indicators are applicable only to insulators and semimetals in which a fixed number of valence bands exist below the Fermi level at every high-symmetry momentum. If one wants to straightforwardly apply this method to SCs, one must examine the representations in the band structure of the Bogoliubov--de Gennes (BdG) Hamiltonian including a gap function.  In fact, this is the approach taken in Ref.~\cite{PhysRevB.98.115150} that recently extended the symmetry indicators to  the 10 Altland-Zirnbauer symmetry classes~\cite{1367-2630-12-6-065010,doi:10.1063/1.3149495,PhysRevB.78.195125}. However, this is not ideal because such a band structure is not available in the standard DFT calculation. Furthermore, in this way, the total number of bands that have to be taken into account can be huge unless one uses an effective tight-binding model.

In this work, we further develop the theory of symmetry indicators exclusively designed for weak-coupling SCs. It enables us to determine the topology of  SCs based on the representations of a finite number of bands below the Fermi surface in the normal phase, although one still has to assume a symmetry transformation property of the gap function.  This is a generalization of the famous criterion~\cite{PhysRevB.81.134508,PhysRevLett.105.097001,PhysRevB.81.220504} that an odd-parity SC with the inversion symmetry is topological when the number of connected Fermi surfaces is odd.  Our refined criterion finds that an odd-parity SC can be topological even when the number of Fermi surfaces is even as we demonstrate in Fig.~\ref{fig2} below using a concrete model.  We also apply our formulae to DFT band structures of several realistic materials to confirm the usefulness of symmetry indicators in the theoretical and experimental search of topological SCs.

\section{Symmetry indicators topological superconductors}
\subsection{Symmetry of Bogoliubov--de Gennes Hamiltonian}
Our discussion in this work is based on the BdG Hamiltonian with the particle-hole symmetry (PHS) $\Xi= \tau_x K$:
\begin{equation}
\label{HBdG}
H_{\bm{k}}^{\text{BdG}} = \begin{pmatrix} H_{\bm{k}} & \Delta_{\bm{k}} \\ \Delta_{\bm{k}}^{\dagger}& -H_{-\bm{k}}^* \end{pmatrix},
\end{equation}
where the gap function $\Delta_{\bm{k}}$ satisfies $\Delta_{\bm{k}}^t=-\Delta_{-\bm{k}}$. The BdG Hamiltonian describes the band structure of the superconducting phase, while $H_{\bm{k}}$ encodes the band structure in the normal phase. We assume a band gap around $E=0$ in the superconducting phase at least at every high-symmetry momentum.  To simplify the analysis, we also set the Fermi level $E_F$ in the normal phase to be $0$. 

Let us recall the spatial symmetry of $H_{\bm{k}}^{\text{BdG}}$.  Let $U_{\bm{k}}(g)$ be a unitary matrix representing an element $g$ of a space group $G$ in the normal phase, satisfying $U_{\bm{k}}(g)H_{\bm{k}}U_{\bm{k}}(g)^{\dagger} = H_{g\bm{k}}$.  When the gap function $\Delta_{\bm{k}}$ obeys  the condition $U_{\bm{k}}(g)\Delta_{\bm{k}}U_{-\bm{k}}(g)^{t} =\chi_g \Delta_{g\bm{k}}$, the BdG Hamiltonian satisfies $U_{\bm{k}}^{\text{BdG}}(g)H_{\bm{k}}^{\text{BdG}}U_{\bm{k}}^{\text{BdG}}(g)^{\dagger} = H_{g\bm{k}}^{\text{BdG}}$ with
\begin{equation}
\label{UBdG}
U_{\bm{k}}^{\text{BdG}}(g)=\begin{pmatrix} U_{\bm{k}}(g) & 0 \\ 0 & \chi_{g}U_{-\bm{k}}(g)^{*} \end{pmatrix}.
\end{equation}
The $U(1)$ phase $\chi_g$ must form a linear representation of $G$ that characterizes the symmetry of the gap function $\Delta_{\bm{k}}$. If the time-reversal symmetry (TRS) is unbroken in the superconducting phase, $\chi_g$ must be either $\pm1$ for all $g\in G$. 


\begin{figure}[t]
\includegraphics[width=1.0\columnwidth]{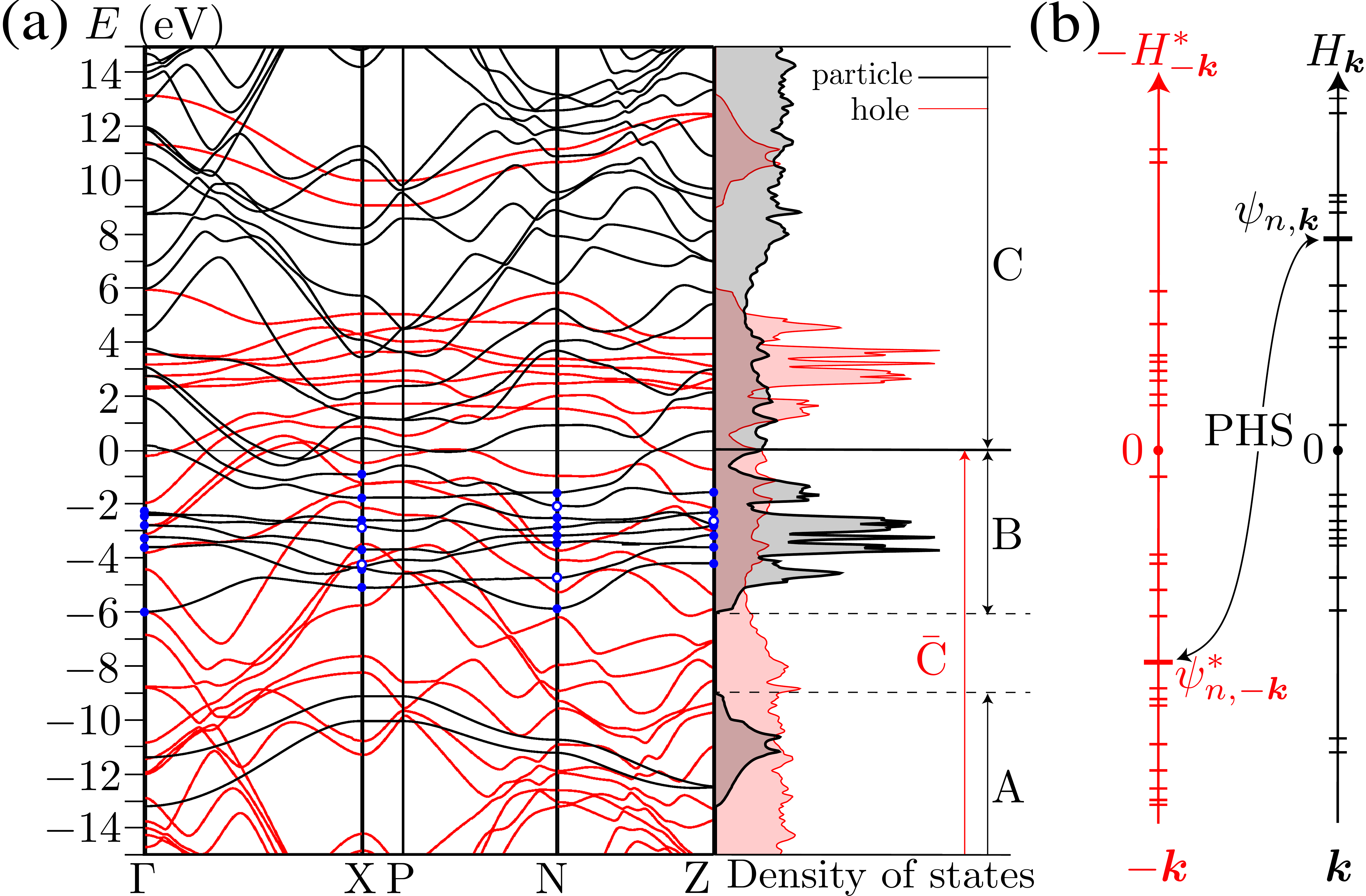}
\caption{\label{fig1} (a) The band structure and density of states of $\beta$-PdBi$_2$ from DFT.  The black (red) curves represent the band structure and the density of states in the particle (hole) sector.  The band structure and the density of states of the hole sector are taken by folding back the particle sector at each $\bm{k}$ as illustrated in (b).
  }
\end{figure}


\subsection{Main results} Let us begin by reviewing the process of computing symmetry indicators in the normal phase using the example of band structure in Fig~\ref{fig1}.  Suppose that the band structure described by $H_{\bm{k}}$ [shown by black curves in Fig~\ref{fig1}(a)] have a band gap around the Fermi energy $E=0$ at every high-symmetry momenta. (There can be band crossing at generic momenta.) We count the number of occurrence  $n_{\bm{k}}^\alpha$ of $u_{\bm{k}}^{\alpha}(g)$ in the finite number of bands below $E=0$ [i.e., in the energy window marked ``$\text{B}$" or ``$\text{A}+\text{B}$" in Fig~\ref{fig1}(a)] and use that data to compute  symmetry indicators of the corresponding insulator or semimetal.  Here, $u_{\bm{k}}^{\alpha}(g)$ ($\alpha=1,2,\ldots$) are irreducible representations of the little group $G_{\bm{k}}$ at a high-symmetry momentum $\bm{k}$. The bands in the window ``$\text{A}$" represents a fully occupied atomic insulator, whose topology is always trivial and can  be neglected from the calculation.

In the superconducting phase, on the other hand, the data required for computing symmetry indicators is the number of representations $(n_{\bm{k}}^\alpha)^{\text{BdG}}$ in the quasi-particle spectrum described by $H_{\bm{k}}^{\text{BdG}}$ below $E=0$~\cite{PhysRevB.98.115150}.  They are basically the combination of ``$\text{A}+\text{B}$" of the particle bands ($H_{\bm{k}}$) and ``$\bar{\text{C}}$" of the hole band ($-H_{-\bm{k}}^*$), the latter of which is shown by red curves in Fig~\ref{fig1}(a). The superconducting gap $\Delta_{\bm{k}}$ modifies these band structures near the Fermi surface. When the matrix size of $H_{\bm{k}}$ is $N$, the total number of bands below $E=0$ considered in this calculation is also $N$. 
Here we face a difficulty for our purpose enabling automated search for 
topological superconductors by combining with DFT calculations. 
This is because $N$ can be arbitrary large due to the existence of irrelevant high-energy bands far above the Fermi level in the normal phase. Physically, we expect that these high-energy bands, as well as the inner bands (``$\text{A}$"), do not affect the nature of the superconductivity. Thus it is customary to derive an effective tight-binding model describing only all relevant bands near the Fermi level \emph{for each material}. In this approach, the matrix size $N$ is kept finite and one can simply apply the original calculation scheme of the symmetry indicators using the BdG Hamiltonian with an assumed gap function $\Delta_{\bm{k}}$.  However, such an approach is not well-suited for automated comprehensive screening of an exhaustive list of materials on database. Note that the use of tight-biding model as an intermediate step is mandatory in this scheme, because otherwise one would introduce a cut-off to the DFT calculation at an arbitrary energy level.  Then the neglected bands would not be fully isolated from those taken into account and the result may be incorrect.  For example, if we set the energy-cutoff to be $3$ eV, $7$ eV, and $15$ eV, we get $\kappa_{1}^{\text{BdG}} =2$, $1$, and $7/2$ (mod $4$), respectively,  using the formula Eq.~\eqref{der} in Appendix.~\ref{App_A}. However, the correct value for this material is  $\kappa_{1}^{\text{BdG}} =3$ (mod $4$).



To avoid this difficulty we develop an alternative approach that does not use a tight-binding model but yet deals with only a finite number of bands. 
This is enabled by the combination of two observations. First we introduce ``weak-pairing assumption" following~\cite{PhysRevB.81.134508, PhysRevLett.105.097001,PhysRevB.81.220504,1701.01944, PhysRevB.99.125149}.
It states that $(n_{\bm{k}}^\alpha)^{\text{BdG}}$ in the superconducting phase does not change even if the limit $\Delta_{\bm{k}}\rightarrow0$ is taken. (This assumption is usually valid; to our knowledge, there are no exceptions.) In this limit, eigenstates of $H_{\bm{k}}^{\text{BdG}}$ and their representations can be exactly deduced from those of $H_{\bm{k}}$ and $-H_{-\bm{k}}^{*}$. 
Let $\psi_{n,\bm{k}}$ be an eigenstate of $H_{\bm{k}}$ with the energy $\epsilon_{n,\bm{k}}$  belonging to the representation $u_{\bm{k}}^{\alpha}(g)$ of $G_{\bm{k}}$. Then, $\psi_{n,-\bm{k}}^*$ is an eigenstate of $-H_{-\bm{k}}^*$ with the energy $-\epsilon_{n,-\bm{k}}$  that belongs to the representation $u_{\bm{k}}^{f_{\bm{k}}(\alpha)}(g)\equiv\chi_g[u_{-\bm{k}}^{\alpha}(g)]^*$ of $G_{\bm{k}}$. This defines an one-to-one map $f_{\bm{k}}$ among irreducible representations of $G_{\pm\bm{k}}$, which can be inverted as $u_{\bm{k}}^{\alpha}(g)=\chi_g[u_{-\bm{k}}^{f_{-\bm{k}}(\alpha)}(g)]^*$. As a result, $(n_{\bm{k}}^\alpha)^{\text{BdG}}$ of the superconducting phase can be expressed in terms of $n_{\bm{k}}^\alpha$ of the normal phase as
\begin{align}
(n_{\bm{k}}^\alpha)^{\text{BdG}}&=n_{\bm{k}}^\alpha\big|_{\text{occ}}+n_{-\bm{k}}^{f_{-\bm{k}}(\alpha)}\big|_{\text{unocc}}. \notag\\
&=(n_{\bm{k}}^\alpha-n_{-\bm{k}}^{f_{-\bm{k}}(\alpha)})\big|_{\text{occ.}}+n_{-\bm{k}}^{f_{-\bm{k}}(\alpha)}\big|_{\text{all bands}}.\label{weak-pair}
\end{align}
Here, the subscript $\text{occ}$ ($\text{unocc}$) refers to the band structure in the normal phase below (above) the Fermi level $E=0$.  Now we utilize our second observation that a band insulator which completely fills all bands is always topologically trivial. Thus we can drop the last term in Eq.~\eqref{weak-pair} as well as the contribution from the inner bands [``$\text{A}$" in Fig.\ref{fig1} (a)] as far as the symmetry indicators are concerned. After all, we get
\begin{equation}
(n_{\bm{k}}^\alpha)^{\text{BdG}}\simeq(n_{\bm{k}}^\alpha-n_{-\bm{k}}^{f_{-\bm{k}}(\alpha)})\big|_{\text{occ}}.
\label{mainresult}
\end{equation}
This is the main theoretical result of this work, which enables us to compute $(n_{\bm{k}}^\alpha)^{\text{BdG}}$ of the superconducting phase solely by the  relevant occupied bands of the normal phase, which is the region ``$\text{B}$" in Fig.\ref{fig1} (a). Note that we are dealing with a metallic band structure and $n_{\bm{k}}^\alpha$ here by itself does not necessarily satisfy the compatibility relations unlike $(n_{\bm{k}}^\alpha)^{\text{BdG}}$.


Below we translate the general formula in \eqref{mainresult} into more convenient forms in applications. Some derivations are included in Appendix~\ref{App_A}.

\begin{figure}
	\includegraphics[width=0.99\columnwidth]{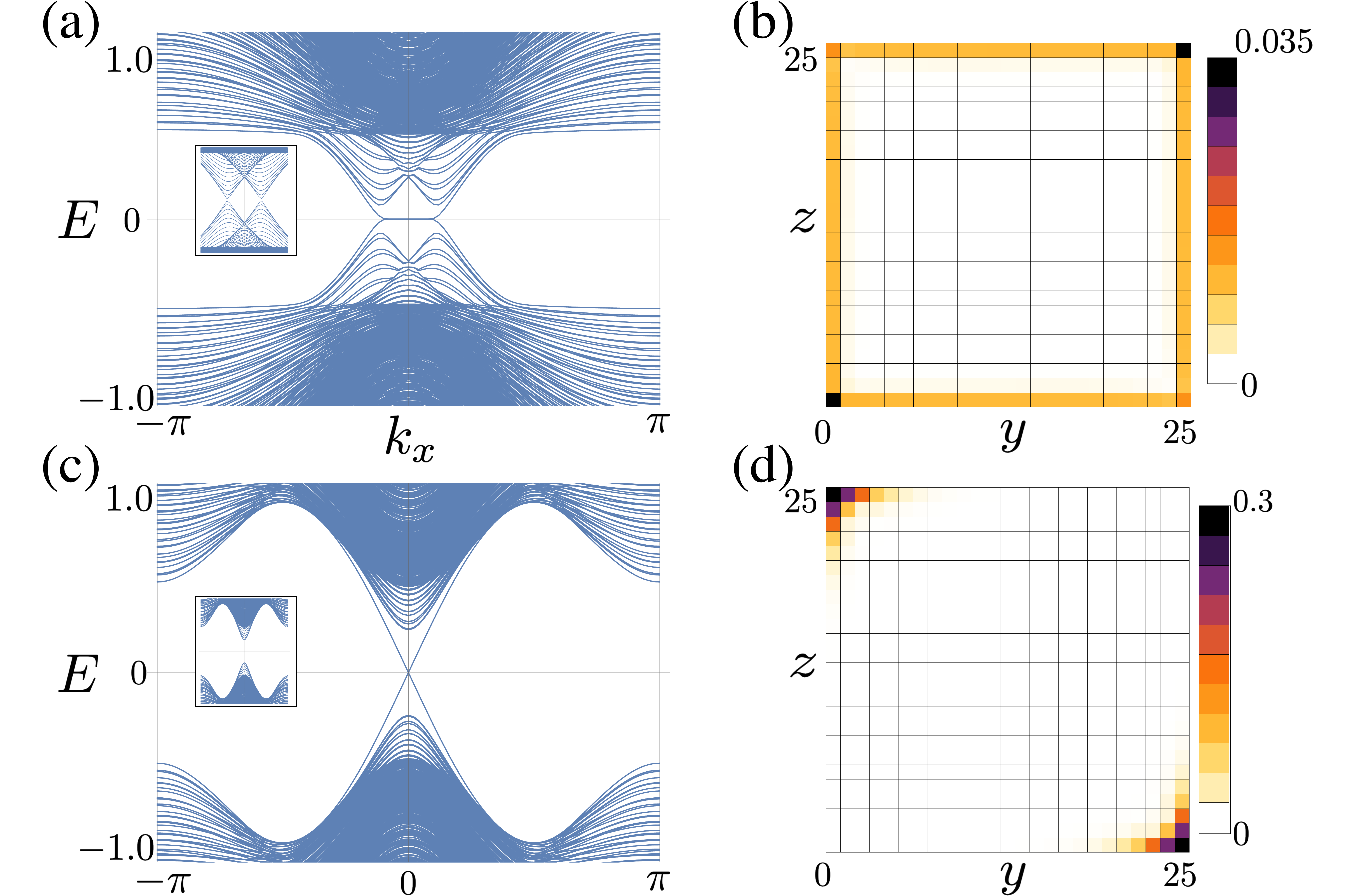} 
	\caption{\label{fig2} 
	Numerical results for $H_{\bm{k}}^{\text{BdG}}$ defined by Eq.~\eqref{He3}. The open boundary condition is imposed on $y$ and $z$ directions with 25 $\times$ 25 unit cells, while the periodic boundary condition is assumed in $x$ direction. 
	(a) The band structure for the choice of gap function $\Delta_{\bm{k}}^{(+,+)}$. 
	(b) The real-space density profile of the zero mode at one of the two Dirac points in (a). The density profiles of zero modes at other momena are listed in the Appendix~\ref{App_B}. 
	(c) The band structure for the choice of gap function $\Delta_{\bm{k}}^{(+,-)}$. 
	(d) The real-space density profile of the zero mode with $k_x =0 $ in (c).
	The insets in (a),(c) are the band structure under the periodic boundary condition in $x$ and $y$ (different values of $k_y$ are superposed.) }
\end{figure}

\subsection{Inversion}
We start with the inversion symmetry $I$.  For odd-parity SCs with TRS (i.e., $\chi_I=-1$ in class DIII), the $\mathbb{Z}_2$ weak indices $\nu_i^{\text{BdG}}$ ($i=1,2,3$)~\cite{PhysRevB.76.045302} and the $\mathbb{Z}_4$ strong index $\kappa_1^{\text{BdG}}$ ~\cite{PhysRevX.8.031070,QuantitativeMappings} can be computed as
\begin{align}
\nu_i^{\text{BdG}}&\equiv\tfrac{1}{4}\sum_{\bm{k}\in\text{2D TRIMs}}\sum_{\alpha=\pm1}\alpha (n_{\bm{k}}^\alpha)^{\text{BdG}} 
\simeq 2\tilde{\nu}_i\in\mathbb{Z},\\
\kappa_1^{\text{BdG}}&\equiv\tfrac{1}{4}\sum_{\bm{k}\in\text{3D TRIMs}}\sum_{\alpha=\pm1}\alpha (n_{\bm{k}}^\alpha)^{\text{BdG}} 
\simeq 2\tilde{\kappa}_1\in\mathbb{Z},\label{inversionindicator}
\end{align}
where $\tilde{\nu}_i\equiv\tfrac{1}{4}\sum_{\bm{k}\in\text{2D TRIMs}}\sum_{\alpha=\pm1}\alpha n_{\bm{k}}^\alpha$ is the sum of the inversion parities of occupied bands over the four appropriate TRIMs (divided by four) and $\tilde{\kappa}_1\equiv\frac{1}{4}\sum_{\bm{k}\in\text{3D TRIMs}}\sum_{\alpha=\pm1}\alpha n_{\bm{k}}^\alpha$ is the same but over all eight TRIMs (divided by four). Most importantly, $\tilde{\nu}_i$ and $\tilde{\kappa}_1$ can be easily computed once the band structure of the normal phase is given. Note that $\tilde{\nu}_i$ and $\tilde{\kappa}_1$ here can be a half-integer, since the band structure in the normal phase is allowed to be metallic.

Let us discuss the nature of topological SCs indicated by $ \kappa_1^{\text{BdG}}$ in Eq.~\eqref{inversionindicator}.  First of all, the parity of $\kappa_1^{\text{BdG}}$ agrees with the 3D winding number $W$ modulo 2.  When the number of connected Fermi surfaces is odd, $\kappa_{1}^\text{BdG}$, and hence $W$, must be odd. This is consistent with the previous studies~\cite{PhysRevB.81.134508,PhysRevLett.105.097001,PhysRevB.81.220504}.   More interesting scenario is when $\kappa_1^{\text{BdG}}=2$ (mod 4) while all weak indices vanishes.  Although this case has been classified to the trivial category in the existing criterion~\cite{PhysRevB.81.134508,PhysRevLett.105.097001,PhysRevB.81.220504}, it still exhibits a nontrivial, possibly higher-order topology as we see below through an example.

As a demonstration, we introduce a toy lattice model of $^3$He B-phase, given by $H_{\bm{k}}^{\text{BdG}}$ in \eqref{HBdG} with 
\begin{align}
H_{\bm{k}}&= \left[t(3-\cos k_x - \cos k_y - \cos k_z)- \mu\right]\sigma_0,\label{He1}\\
\Delta_{\bm{k}}^{(\xi)}&= -\xi\Delta\left(\sin k_x \sigma_x + \sin k_y \sigma_y+\sin k_z \sigma_z \right)i\sigma_y,\label{He2}
\end{align}
where $\sigma_j$ is the Pauli matrix.  Below we set $t=\mu=\Delta=1$. 
The model has the inversion symmetry $U_{\bm{k}}(I)=\sigma_0$ and the TRS $U_{\mathcal{T}}=-i\sigma_y$.
Only the $\Gamma$ point is occupied by the two even-parity bands.  We thus get $\tilde{\kappa}_1=1/2$ and $\kappa_{1}^{\text{BdG}}=1$, which implies that $W$ is odd.
We indeed find $W = \pm 1$ depending on $\xi = \pm 1$.

To realize the case with $\kappa_{1}^{\text{BdG}}=2$, let us take two copies of this model:
\begin{equation}
H_{\bm{k}}'=
\begin{pmatrix}
H_{\bm{k}}&V\\
V^\dagger&H_{\bm{k}}
\end{pmatrix},\quad \Delta_{\bm{k}}^{(\xi_1,\xi_2)}=
\begin{pmatrix}
\Delta_{\bm{k}}^{(\xi_1)}&0\\
0&\Delta_{\bm{k}}^{(\xi_2)}
\end{pmatrix},\label{He3}
\end{equation}
where $V=-i\bm{m}\cdot\bm{\sigma}$ ($|\bm{m}|<1$) represents a perturbation respecting both the inversion and TRS and we set $\bm{m}=\frac{1}{2}(0,\frac{1}{\sqrt{2}},\frac{1}{\sqrt{2}})$. $H_{\bm{k}}'$ has four bands occupying the $\Gamma$ point and we get $\tilde{\kappa}_1=\tilde{\nu}_i=1$ so that $\kappa_1^{\text{BdG}}= 2$ (mod 4) and $\nu_i^{\text{BdG}}=0$ (mod 2) for $i=1,2,3$. 
When we choose $\Delta_{\bm{k}}^{(+,-)}$ for the gap function, the winding $W=0$ and the corresponding $H_{\bm{k}}^{\text{BdG}}$ realizes a higher-order topological SC with 1D helical Majorana modes as illustrated in Fig.~\ref{fig2} (c,d)~\cite{PhysRevB.97.205136}. On the other hand, when we assume $\Delta_{\bm{k}}^{(+,+)}$ instead,
 the winding $W=2$ implies that the 2D surface is gapless [see Fig.~\ref{fig2} (a,b)].  In fact, this case has co-existing 2D surface modes protected by internal symmetries together with 1D hinge modes protected by time-reversal and inversion symmetry. A similar but distinct hybrid surface state was reported in Ref.~\onlinecite{PhysRevB.99.125149}.


To demonstrate the advantage of our method that does not rely on an effective tight-binding model at any step, let us apply our result directly to the real DFT band structure of $\beta\text{-PdBi}_2$ in Fig.~\ref{fig1}.~\footnote{Our ab initio calculations are performed by using WIEN2K~\cite{wien2k} and all material information is taken from ``Materials Project"~\cite{Jain2013}.}  We assume that (i) the inversion and TRS remain unbroken and that (ii) a full gap~\cite{Iwaya:2017aa} with odd inversion parity ($\chi_I=-1$)~\cite{1810.11265} opens in the superconducting phase. 
In Fig.~\ref{fig1} (a), the solid (open) circles in the energy window B represent positive (negative) inversion parities. According to that, the sum of inversion parities are $12$, $8$, $8$, $10$, respectively, at $\Gamma$, $X$, $N$, $Z$ including the spin degeneracy.  Taking into account the multiplicity of $X$ and $N$ points (i.e., there are two $X$ points and four $N$ points), we get $\tilde{\kappa}_1 = \tfrac{1}{4}(12+ 2\times 8 + 4 \times 8 + 10 )= 35/2$ and $\kappa_1^{\text{BdG}}=2\tilde{\kappa}_1 = 3$ mod 4.
This implies that the 3D winding number $W$ is nontrivial, which is consistent with the observation of topological surface states~\cite{Sakano:2015aa,LV2017852}.

\begin{figure}[t]
	\includegraphics[width=0.99\columnwidth]{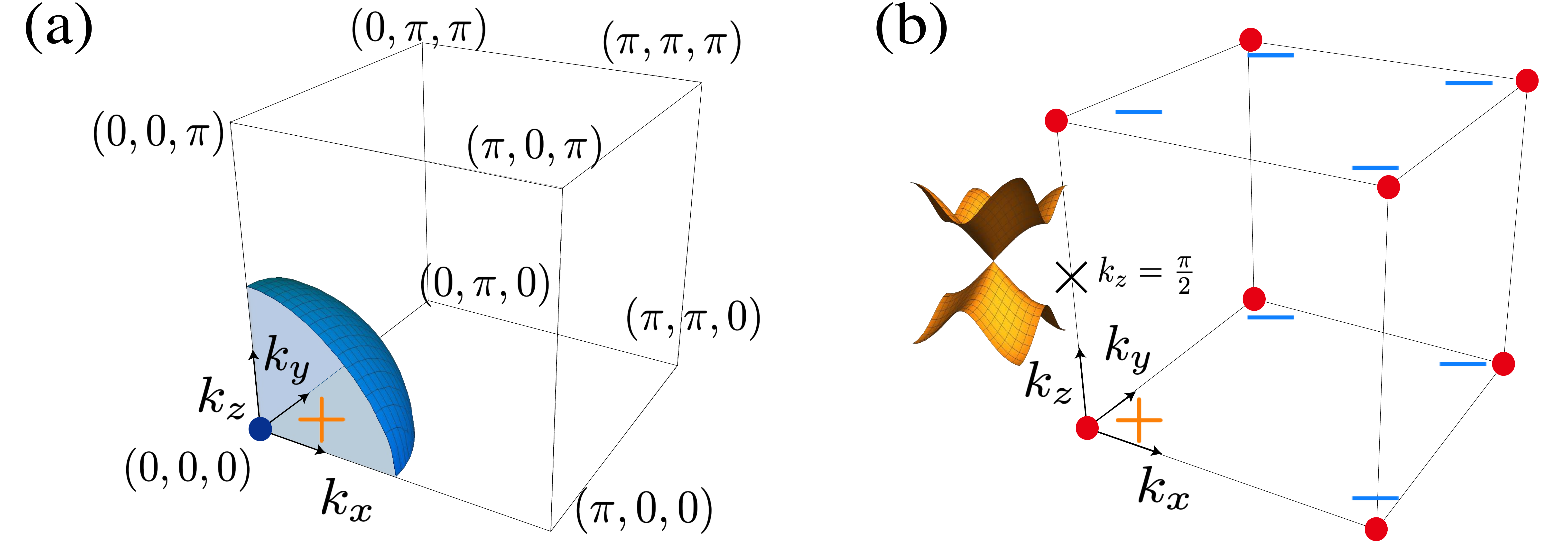} 
	\caption{\label{fig3} 
		The 3D chiral $p$-wave SCs described by Eqs.~\eqref{ChP1} and \eqref{ChP2}. (a) The Fermi surface and the inversion parity of the normal phase. (b) The inversion parity and Weyl nodes in the superconducting phase.
}\end{figure}

\subsection{Nodal SCs}
Although our focus in this work is mainly on fully gapped SCs, our theory can also be equally applied to nodal SCs as far as the nodes do not locate at high-symmetry points in the Brillouin zone. For example, let us again discuss the sum of inversion parities over eight TRIMs but this time without assuming TRS (class D)~\cite{PhysRevB.98.115150}:
\begin{align}
\mu_1^{\text{BdG}} \equiv\tfrac{1}{2}\sum_{\bm{k}\in\text{3D TRIMs}}\sum_{\alpha=\pm1}\alpha (n_{\bm{k}}^\alpha)^{\text{BdG}}\in\mathbb{Z}.\label{inv2}
\end{align}
When $\mu_1^{\text{BdG}}$ is odd, the Chern numbers on $k_{z}=0$ plane and $k_{z}=\pi$ plane are different and there must be some nodes in the quasi-particle spectrum between these planes.  
As an example, let us consider a 3D extension of the chiral $p$-wave SC:
\begin{align}
H_{\bm{k}}&= t(3-\cos k_x - \cos k_y - \cos k_z)- \mu,\label{ChP1}\\
\Delta_{\bm{k}}&=\Delta(\sin k_x +i \sin k_y).\label{ChP2}
\end{align}
The single band in the normal phase occupies only the $\Gamma$ point  and 
$\mu_1^{\text{BdG}}=\pm1$ is odd.  Indeed, there is a pair of Weyl points at $\bm{k}=\pm(0,0,\pi/2)$ as illustrated in Fig.~\ref{fig3} (b). This is the superconducting generalization of the symmetry-enforced Weyl semimetal discussed in Ref.~\onlinecite{PhysRevB.85.165120}. 



\subsection{Rotation}
Next, let us discuss formulae diagnosing the (mirror) Chern numbers based on $n$-fold rotation eigenvalues~\cite{PhysRevB.86.115112,1701.01944}.  We summarize our results in Tables~\ref{tab:Chern}, \ref{tab:RD}, which enable us to determine the (mirror) Chern numbers of SCs modulo $n$ using the rotation eigenvalues in the normal phase. 
There are additional constraints on mirror Chern numbers, such as $C_{+i}=C_{-i}$ when $\chi_{M_{xy}}=+1$ and $C_{+i}=-C_{-i}$ when TRS is unbroken in the superconducting phase.  If representations are not consistent with them, the gap $|\Delta_{\bm{k}}|$ must vanish at some $\bm{k}$ resulting in a nodal SC.

The simplest example is given by the $k_z=0$ plane of the chiral $p$-wave SC in Eq.~\eqref{ChP2}. The model has $C_4$-rotation symmetry with $U_{\bm{k}}(C_4)=1$ and $\chi_{C_4}=i$.  Recalling that there is only one band and it occupies only $\Gamma$, we apply the formula for $n=4$ as $R=1\times 1/1=1$ and $\Delta=2\times 0 -1-0=-1$. Hence, we find $e^{\frac{2\pi i}{4}C}=i^{-1}\times 1^2=-i$. This agrees with the actual value of $C=-1$.


As a more nontrivial demonstration, let us apply our results to $\text{Sr}_2\text{Ru}\text{O}_4$.  There are many proposals of the specific form of the gap function for this material~\cite{doi:10.1143/JPSJ.81.011009,PhysRevLett.111.087002,PhysRevLett.121.157002,PhysRevX.7.011032}. Here we consider the two possibilities studied in Ref.~\onlinecite{PhysRevLett.111.087002}.  The space group of the two superconducting phases is $I4/m$ because mirror symmetries and the TRS are broken by cooper pairs.

Firstly, we discuss the tight-binding model used in Fig.~\ref{fig4}.  We present the details of the model and the symmetry representations in Appendix~\ref{App_C}.  It has a $C_4$-rotation symmetry, a mirror symmetry $M_{xy}$, and TRS. Based on the band structure in Fig.~\ref{fig4}, we get $R_{+i}=(R_{-i})^*=e^{\frac{(-3)+1+(-3)}{4}\pi i}/i=e^{\frac{\pi i}{4}}$ and  $\Delta_{+i}=\Delta_{-i}=2\times 1-3-0=-1$. Hence, if we set $\chi_{M_{xy}}=+1$ and $\chi_{C_4}=-i$, we get $C_{+i}=C_{-i}=+1$ (mod 4). If we use $\chi_{M_{xy}}=-1$ and $\chi_{C_4}=+1$ instead, we get $C_{+i}=-C_{-i}=+1$ (mod 4). These results are consistent with Ref.~\onlinecite{PhysRevLett.111.087002}. 

\begin{figure}[t]
	\includegraphics[width=1.0\columnwidth]{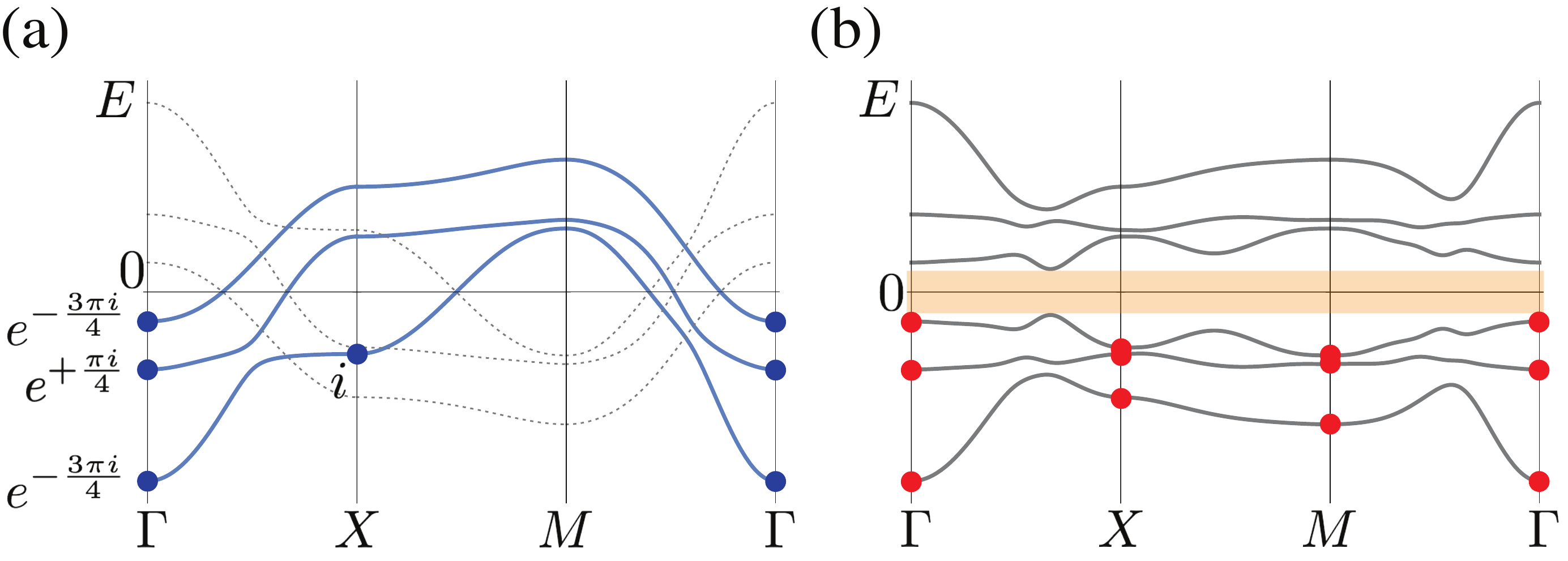}
	\caption{\label{fig4} The band structure of the three-orbital tight-binding model on a 2D square lattice that models a $\text{Ru}\text{O}_2$ plane of $\text{Sr}_2\text{Ru}\text{O}_4$~\cite{Sigrist}. Each band is doubly degenerate because of the inversion symmetry and TRS.  (a) and (b) are for the normal and the superconducting phase, respectively.  Blue dots in (a) and red dots in (b) indicate occupied states at high-symmetry momenta.  Dotted curves in (a) represent the band structure of $-H_{\bm{-k}}^*$, and the numbers aside blue dots are rotation eigenvalues of the $M_{xy}=+i$ sector. (Those for the other sector are given by the complex conjugation.)
}\end{figure}

Lastly, let us apply our formulae to the DFT band strucutre of this material. To avoid complications by the body-centered lattice, we translate all informations of representations into the primitive lattice that contains two primitive unit cells(see Appendix~\ref{App_E}).  
Using formulae in Table~\ref{tab:RD} to the DFT band structure included in Appendix~\ref{App_F}, we get $R_{\sigma} = -\sigma$ and $\Delta_\sigma=-2$ in the $k_z=0$ plane and $R_{\sigma} = -1$ and $\Delta_\sigma=-2$ in the $k_z=\pi$ plane for each mirror sector $\sigma=\pm i$.  Now we use Table~\ref{tab:Chern}.  Assuming $\chi_{M_{xy}}=+1$ and $\chi_{C_4}=-i$ frist, we get $C_{\sigma}=+2$ (mod 4) on both $k_z=0$ and $\pi$.
Using $\chi_{M_{xy}}=-1$ and $\chi_{C_4}=+1$ instead, we get $C_{+i}=-C_{-i}=+2$ (mod 4) at $k_z=0$ and $C_{\sigma}=0$ (mod 4)  at $k_z=\pi$. These results are consistent with the picture of stacked $\text{Ru}\text{O}_2$ layers discussed in Ref.~\onlinecite{PhysRevLett.111.087002}.

\begin{table}
		\caption{\label{tab:Chern}Formulas for diagnosing (mirror) Chern numbers for SCs based on the $n$-fold rotation eigenvalues ($n=2$, $3$, $4$, and $6$). $R$ and $\Delta$ are defined in Table~\ref{tab:RD}. When $M_{xy}$ exists, $R_\sigma$ and $\Delta_\sigma$ for each mirror sector $\sigma=\pm i$ are defined in the same way.  If the normal phase has TRS, $R_{+i}=(R_{-i})^*$ and $\Delta_{+i}=\Delta_{-i}$.}
	\begin{ruledtabular}
		\begin{tabular}{cl}
			Symmetry & $\quad$(mirror) Chern numbers\\
			\hline
			$C_n$ &  $e^{\frac{2\pi i}{n}C} =(\chi_{C_n})^{\Delta}{R}^2$.\\
			$\quad$ $C_n$ \& $M_{xy}$ ($\chi_{M_{xy}}=+1$)$\quad$ & $e^{\frac{2\pi i}{n}C_\sigma} =(\chi_{C_n})^{\Delta_{-\sigma}}R_{+i}R_{-i}$.\\
			$C_n$ \& $M_{xy}$ ($\chi_{M_{xy}}=-1$) & $e^{\frac{2\pi i}{n}C_\sigma} =(\chi_{C_n})^{\Delta_{\sigma}}(R_{\sigma})^2$.\\
		\end{tabular}
	\end{ruledtabular}
\end{table}

\begin{table}
	\caption{\label{tab:RD}Definition of $R$ and $\Delta$ in Table~\ref{tab:Chern}. Here, $\zeta_{\bm{k}}$, $\theta_{\bm{k}}$, $\xi_{\bm{k}}$, and $\eta_{\bm{k}}$ respectively represent the product of $2$, $3$, $4$, and $6$-fold rotation eigenvalues over all occupied bands at $\bm{k}$.  $N_{\bm{k}}$ is the total number of occupied bands at $\bm{k}$.}
	\begin{ruledtabular}
		\begin{tabular}{ccl}
			$\quad n\quad$& $\quad\quad\quad R\quad\quad\quad$ &$\quad\quad \Delta \quad\quad\quad$ \\
			\hline
			$2$ &$\zeta_\Gamma\zeta_X\zeta_Y\zeta_M$& $N_X+N_Y-N_\Gamma-N_M$ \\
			$3$ &$\theta_K\theta_{K'}/(\theta_\Gamma)^2$ & $2N_\Gamma-N_K-N_{K'}$\\
			$4$ &$\xi_\Gamma\xi_M/\zeta_X$ & $2N_X- N_\Gamma - N_M$\\
			$6$ &$\eta_\Gamma\theta_K/\zeta_M$ & $3N_M-N_\Gamma - 2N_K$\\
		\end{tabular}
	\end{ruledtabular}
\end{table}


\subsection{Rotoinversion}
Finally the rotoinversion symmetry $S_4$ also defines a three-dimensional $\mathbb{Z}_2$ strong index $\kappa_4^{\text{BdG}}$ in class DIII~\cite{PhysRevX.8.031070}.  $\kappa_4^{\text{BdG}}$ can be non-trivial when $\chi_{S_4}=-1$. In this case we have
\begin{equation}
\kappa_4^{\text{BdG}}\simeq\tfrac{1}{\sqrt{2}}\sum_{\bm{k}\in K_4}\sum_{\alpha=1,3,5,7}e^{i\frac{\alpha\pi}{4}} n_{\bm{k}}^\alpha\in\mathbb{Z}, \label{rotinversionindicator}
\end{equation}
where $n_{\bm{k}}^\alpha$ represents the number of rotoinversion eigenvalues $e^{i\frac{\alpha\pi}{4}}$ at four $S_4$-symmetric momenta $K_4$.



\section{Conclusion}
In this work, we extended the theory of symmetry indicators for weak-coupling SCs and derived several useful formulae in the search for new topological SCs.  Our general results, such as Eq.~\eqref{inversionindicator} and   Tables~\ref{tab:Chern}, \ref{tab:RD}, enable us to determine the topology of SCs based on the information of representations $n_{\bm{k}}^\alpha$ of occupied bands in the normal phase and the symmetry property $\chi_g$ of the assumed gap function. Conversely, we can use our method to narrow down candidates of the correct gap function of superconductors. Given the information of $n_{\bm{k}}^\alpha$ of the material, there is not much computational cost to calculate the indicators for all possible values of $\chi_g$. One can see how the result fits to the known property of the material and propose 
what the necessary future experiments are to distinguish the symmetry of superconductivity.


In addition to the comprehensive material investigations through DFT calculations~\cite{Tang:2019aa,Vergniory:2019aa,Zhang:2019aa}, the field of materials informatics has been developing rapidly~\cite{Potyrailo:2011aa,1705.01043,1742-6596-699-1-012001} due to the progress of machine learning and used to identify new SCs~\cite{1808.07973,1882-0786-11-9-093101}.  Our symmetry indicators for SCs established in this work can be easily combined with these techniques and should lead to the discovery of many more topological SCs.

\begin{acknowledgments}
	The authors would like to thank M. T. Suzuki, T. Nomoto, M. Hirayama, R. Arita, J. Ishizuka, A. Daido, S. Sumita, Y. Nomura, Y. Tada, and M. Sato for fruitful discussion. S.O. is especially grateful to M. Hayashi and G. Qu for their technical support. The work of S.O. is partially performed during his stay at Kyoto university supported by the exchange program of “Topological Materials Science of MEXT, Japan. The work of S.O. is supported by Materials Education program for the future leaders in Research, Industry, and Technology (MERIT).  
	The work of Y. Y. is supported by JSPS KAKENHI Grant No.~JP15H05884, JP15H05745, JP18H04225, 
	JP18H01178, and JP18H05227. 
	The work of H. W. is supported by JSPS KAKENHI Grant No.~JP17K17678 and by JST PRESTO Grant No.~JPMJPR18LA. 
\end{acknowledgments}

\appendix
\section{Derivations of formulae}
\label{App_A}
Here we present the derivation the formulas of symmetry indicators.  Let $\psi_{n,\bm{k}}$ be an eigenstate of $H_{\bm{k}}$ with the energy $\epsilon_{n,\bm{k}}$  belonging to the representation $u_{\bm{k}}^{\alpha}(g)$ of $G_{\bm{k}}$. 
As explained in the main text, $\psi_{n,-\bm{k}}^*$ is an eigenstate of $-H_{-\bm{k}}^*$ with the energy $-\epsilon_{n,-\bm{k}}$  that belongs to the representation $u_{\bm{k}}^{f_{\bm{k}}(\alpha)}(g)\equiv\chi_g[u_{-\bm{k}}^{\alpha}(g)]^*$ of $G_{\bm{k}}$. 

\subsection{Inversion}
Let us start with Eq.~(5) of the main text.  We assume the inversion symmetry and look at a time-reversal invariant momentum (TRIM) $\bm{k}$.  We denote by $\xi_{n,\bm{k}}$ $(=\pm1)$ the inversion parity of $\psi_{n,\bm{k}}$. Also, let $n_{\bm{k}}^\alpha|_{\text{occ}}^{(\text{particle})}$ ($n_{\bm{k}}^\alpha|_{\text{unocc}}^{(\text{particle})}$) be the number of occupied (unoccupied) states with the inversion parity $\alpha$ in the particle sector $H_{\bm{k}}$. Recall that $\psi_{n,\bm{k}}$ in the particle sector corresponds to $\psi_{n,\bm{k}}^{*}$ in the hole sector $-H_{\bm{k}}^*$, which has the inversion parity $\chi_I \xi_{n,\bm{k}}$. Namely, the parity of the particle and the hole sector flips sign when $\chi_I=-1$. Therefore, the number of occupied states of $-H_{\bm{k}}^*$ with the inversion parity $\alpha$ is given by
\begin{equation}
n_{\bm{k}}^\alpha|_{\text{occ}}^{(\text{hole})}=n_{\bm{k}}^{\chi_I\alpha}|_{\text{unocc}}^{(\text{particle})}.
\end{equation}
Combining the particle and the hole contributions, we get
\begin{align}
\kappa_{1}^{\mathrm{BdG}} &=\tfrac{1}{4}\sum_{\bm{k} \in \mathrm{3D TRIMs}}\sum_{\alpha = \pm 1} \alpha (n_{\bm{k}}^{\alpha})^{\text{BdG}}\nonumber\\
&= \tfrac{1}{4}\sum_{\bm{k} \in \mathrm{3D TRIMs}}\sum_{\alpha = \pm 1} \alpha(n_{\bm{k}}^\alpha|_{\text{occ}}^{(\text{particle})}+n_{\bm{k}}^\alpha|_{\text{occ}}^{(\text{hole})})\nonumber\\
\label{der}
&= \tfrac{1}{4}\sum_{\bm{k} \in \mathrm{3D TRIMs}}\sum_{\alpha = \pm 1} \alpha(n_{\bm{k}}^{\alpha}|_{\text{occ}}^{(\text{particle})}+n_{\bm{k}}^{\chi_I\alpha}|_{\text{unocc}}^{(\text{particle})}).
\end{align}
To translate the information of unoccupied bands into that of occupied bands, we use the fact that the system is always topologically trivial when all bands are occupied. In other words,
\begin{align}
\label{trivial}
0 &= \tfrac{1}{4}\sum_{\bm{k} \in \mathrm{3D TRIMs}}\sum_{\alpha = \pm 1} \alpha(n_{\bm{k}}^{\alpha}|_{\text{occ}}^{(\text{particle})} +n_{\bm{k}}^{\alpha}|_{\text{unocc}}^{(\text{particle})}). \ \  (\text{mod}\ 4)
\end{align}
From Eq. \eqref{der} and \eqref{trivial}, the formula for $\kappa_{1}^{\text{BdG}}$ in the main text can be derived. 
\begin{align}
\kappa_{1}^{\text{BdG}} &= \begin{cases} 0 \ \ (\chi_I = +1) \\ \frac{1}{2}\displaystyle{\sum_{\bm{k}\in \mathrm{3D TRIMs}}}\displaystyle{\sum_{\alpha = \pm 1}} \alpha n_{\bm{k}}^{\alpha}|_{\text{occ}}^{(\text{particle})}  \ \ (\chi_I = -1)\end{cases}  \ \ (\text{mod}\ 4)
\end{align}
In Eqs. (5) and (6) of the main text, we omit labels ``(particle)" and ``occ".  The derivation of $\nu_i$ in Eq.~(5) is identical.


\subsection{Mirror Chern number}
Next we discuss the results summarized in Table I and II of the main text.  We consider a 2D system with the $n$-fold rotation symmetry about the $z$ axis and the mirror symmetry about the $xy$ plane.  We assume
\begin{align}
\label{rotation-parity-Cn}
&U^{\text{BdG}}_{\bm{k}}(C_n) = \begin{pmatrix} U_{\bm{k}}(C_n)& 0 \\ 0 & \chi_{C_n}U_{\bm{k}}(C_n)^{*}\end{pmatrix},\\
\label{mirror-parity-Cnh}
&U^{\text{BdG}}_{\bm{k}}(M_{xy}) = \begin{pmatrix} U_{\bm{k}}(M_{xy})  & 0 \\ 0 & \chi_{M_{xy}}U_{\bm{k}}(M_{xy})^{*}\end{pmatrix}
\end{align}
with ${\chi_{M_{xy}}}^2 = {\chi_{C_n}}^n = 1$.  We can calculate the mirror Chern number modulo $n$ by using the rotation eigenvalues at high-symmetry momenta~\cite{PhysRevB.86.115112} for each mirror sector. 
Suppose that $\bm{k}$ is invariant under the rotation symmetry. Since $H_{\bm{k}}^{\text{BdG}}$, $U^{\text{BdG}}_{\bm{k}}(C_n)$, and $U^{\text{BdG}}_{\bm{k}}(M_{xy})$ all commute, we can simultaneously diagonalize them.  Let us denote by $\psi_{m,\bm{k}}^{\sigma}$ the Bloch function of the $m$-th band with the mirror eigenvalue $\sigma=\pm i$.  If the $n$-fold rotation eigenvalue  of $\psi_{m,\bm{k}}^{\sigma}$ is $\xi_{m,\bm{k}}^{\sigma}$ , the eigenvalues of $(\psi_{m,\bm{k}}^{\sigma})^{*}$ in the mirror sector $\sigma$ is $\chi_{C_n} (\xi_{m,\bm{k}}^{-\chi_{M}\sigma})^{*}$. 



Below we focus on the case of $n=4$. The derivation for other cases can be done in the same way. Let us denote by $\xi_{\bm{k}}^{\sigma} |_{\text{occ}}^{(\text{particle})}$  the product of four-fold rotation eigenvalues of occupied bands in the particle sector $H_{\bm{k}}$ at $\Gamma=(0,0)$ or $M=(\pi,\pi)$. Similarly, let $\zeta_X^{\sigma} |_{\text{occ}}^{(\text{particle})}$ be the product of two-fold rotation eigenvalues of occupied bands in the particle sector $H_{X}$ at $X=(\pi,0)$.
We define the same for unoccupied bands analogously. Because the topology of all bands in total is always trivial, we have
\begin{align}
\label{trivial_Cn}
1 &= \left(\frac{\xi_{\Gamma}^{\sigma}\xi_{M}^{\sigma}}{\zeta_{ X}^{\sigma}}\Big|_{\text{occ}}^{(\text{particle})}\right) \left(\frac{\xi_{\Gamma}^{\sigma}\xi_{M}^{\sigma}}{\zeta_{ X}^{\sigma}}\Big|_{\text{unocc}}^{(\text{particle})}\right).
\end{align}

Recall that the products $\xi^{\sigma} |_{\text{unocc}}^{(\text{particle})}$ and $\zeta^{\sigma}|_{\text{unocc}}^{(\text{particle})}$ correspond to those in the hole sector $-H_{\bm{k}}^*$ as
\begin{align}
\xi_{\bm{k}}^{\sigma} |_{\text{occ}}^{(\text{hole})} &= (\chi_{C_4})^{N_{\bm{k}}^{-\chi_{M_{xy}}\sigma}|_{\text{unocc}}^{(\text{particle})}}\left(\xi_{\bm{k}}^{-\chi_{M_{xy}}\sigma} |_{\text{unocc}}^{(\text{particle})}\right)^{*},\\
\zeta_{\bm{k}}^{\sigma} |_{\text{occ}}^{(\text{hole})} &= (\chi_{C_4})^{2N_{\bm{k}}^{-\chi_{M_{xy}}\sigma}|_{\text{unocc}}^{(\text{particle})}}\left(\zeta_{\bm{k}}^{-\chi_{M_{xy}}\sigma} |_{\text{unocc}}^{(\text{particle})}\right)^{*},
\end{align}
where $N_{\bm{k}}^{\sigma}|_{\text{unocc}}^{(\text{particle})}$ represents the number of unoccupied bands of $H_{\bm{k}}$ in the mirror sector $\sigma$. Then,
\begin{widetext}
	\begin{align}
	e^{i\tfrac{2\pi}{4}C^{\sigma}} &= \frac{(\xi_{\Gamma}^{\sigma})^{\text{BdG}}(\xi_{ M}^{\sigma})^{\text{BdG}}}{(\zeta_{X}^{\sigma})^{\text{BdG}}} = 
	\left(\frac{\xi_{\Gamma}^{\sigma}\xi_{M}^{\sigma}}{\zeta_{ X}^{\sigma}}\Big|_{\text{occ}}^{(\text{particle})}\right) 
	\left(\frac{\xi_{\Gamma}^{\sigma}\xi_{M}^{\sigma}}{\zeta_{ X}^{\sigma}}\Big|_{\text{occ}}^{(\text{hole})}\right) \notag
	\\
	&= (\chi_{C_4})^{\big(N_{\Gamma}^{-\chi_{M_{xy}}\sigma}+N_{M}^{-\chi_{M_{xy}}\sigma}-2N_{X}^{-\chi_{M_{xy}}\sigma}\big)\big|_{\text{unocc}}^{(\text{particle})}}
	\left(\frac{\xi_{\Gamma}^{\sigma}\xi_{M}^{\sigma}}{\zeta_{X}^{\sigma}}\Big|_{\text{occ}}^{(\text{particle})}\right)
	\left(\frac{\xi_{\Gamma}^{-\chi_{M_{xy}}\sigma}\xi_{M}^{-\chi_{M_{xy}}\sigma}}{\zeta_{X}^{-\chi_{M_{xy}}\sigma}}\Big|_{\text{unocc}}^{(\text{particle})}\right)^{*}.
	\end{align}

Substituting Eq.~\eqref{trivial_Cn} for this equation, 
	\begin{align}
	\label{third-line}
	e^{i\tfrac{2\pi}{4}C^{\sigma}}
	&=  (\chi_{C_4})^{\big(N_{\Gamma}^{-\chi_{M_{xy}}\sigma}+N_{M}^{-\chi_{M_{xy}}\sigma}-2N_{X}^{-\chi_{M_{xy}}\sigma}\big)\big|_{\text{unocc}}^{(\text{particle})}}
	\left(\frac{\xi_{\Gamma}^{\sigma}\xi_{M}^{\sigma}}{\zeta_{X}^{\sigma}}\Big|_{\text{occ}}^{(\text{particle})}\right)
	\left(\frac{\xi_{\Gamma}^{-\chi_{M_{xy}}\sigma}\xi_{M}^{-\chi_{M_{xy}}\sigma}}{\zeta_{X}^{-\chi_{M_{xy}}\sigma}}\Big|_{\text{occ}}^{(\text{particle})}\right).
	\end{align}

Finally we rewrite $N_{\bm{k}}^{\sigma}|_{\text{unocc}}^{(\text{particle})}$ in the exponent by the number of occupied bands $N_{\bm{k}}^{\sigma}|_{\text{occ}}^{(\text{particle})}$ at $\bm{k}$. Since the total number of bands $N^{\sigma}|_{\text{tot}}^{(\text{particle})}=N_{\bm{k}}^{\sigma}|_{\text{occ}}^{(\text{particle})}+N_{\bm{k}}^{\sigma}|_{\text{unocc}}^{(\text{particle})}$ does not depend on $\bm{k}$, we have
	\begin{equation}
	\big(N_{\Gamma}^{-\chi_{M_{xy}}\sigma}+N_{M}^{-\chi_{M_{xy}}\sigma}-2N_{X}^{-\chi_{M_{xy}}\sigma}\big)\big|_{\text{unocc}}^{(\text{particle})}=-\big(N_{\Gamma}^{-\chi_{M_{xy}}\sigma}+N_{M}^{-\chi_{M_{xy}}\sigma}-2N_{X}^{-\chi_{M_{xy}}\sigma}\big)\big|_{\text{occ}}^{(\text{particle})}.
	\end{equation}

Hence, we get
	\begin{align}	
	e^{i\tfrac{2\pi}{4}C^{\sigma}} &=  (\chi_{C_4})^{\big(2N_{X}^{-\chi_{M_{xy}}\sigma}-N_{\Gamma}^{-\chi_{M_{xy}}\sigma}-N_{M}^{-\chi_{M_{xy}}\sigma}\big)\big|_{\text{occ}}^{(\text{particle})}}
	\left(\frac{\xi_{\Gamma}^{\sigma}\xi_{M}^{\sigma}}{\zeta_{X}^{\sigma}}\Big|_{\text{occ}}^{(\text{particle})}\right)
	\left(\frac{\xi_{\Gamma}^{-\chi_{M_{xy}}\sigma}\xi_{M}^{-\chi_{M_{xy}}\sigma}}{\zeta_{X}^{-\chi_{M_{xy}}\sigma}}\Big|_{\text{occ}}^{(\text{particle})}\right).
	\end{align}
	\end {widetext}
	This reproduces Table I and II of the main text for $n=4$, where we omit labels ``(particle)" and ``occ."
	
	\subsection{Rotoinversion}
	Next, we discuss Eq.(10) in the main text. We consider a system with the four-fold rotoinversion symmetry. Let us assume
	\begin{align}
	\label{rotoinversion-parity-S4}
	U^\text{BdG}_{\bm{k}}(S_4) &= \begin{pmatrix} U_{\bm{k}}(S_4)& 0 \\ 0 & \chi_{s_4}U_{\bm{k}}(S_{4})^{*} \end{pmatrix},
	\end{align}
	where ${\chi_{S_4}} = e^{i\frac{2m}{4}\pi} \ (m=1,2,3,4)$. 
	Suppose that $\bm{k}$ is invariant under the rotoinversion symmetry. When  the rotoinversion eigenvalue of $\psi_{n,\bm{k}}$ is $\xi_{n,\bm{k}}^{\alpha} = e^{i\tfrac{\alpha\pi}{4}}$, the rotoinversion eigenvalue of $(\psi_{n,\bm{k}})^{*}$ is $\chi_{S_4} (\xi_{n,\bm{k}}^{\alpha})^{*}=e^{i(2m-\alpha)\tfrac{\pi}{4}}$. As explained in the ``Inversion" section, we denote the number of occupied (unoccupied) states of $H_{\bm{k}}$ with the rotoinversion eigenvalues $\xi_{n,\bm{k}}^{\alpha}$ 
	by $n_{\bm{k}}^\alpha|_{\text{occ}}^{(\text{particle})}$ $\left(n_{\bm{k}}^\alpha|_{\text{unocc}}^{(\text{particle})}\right)$. Then, the number of occupied states of $-H_{\bm{k}}^*$ with the rotoinversion eigenvalues $\xi_{n,\bm{k}}^{\alpha}$ is given by
	\begin{equation}
	n_{\bm{k}}^\alpha|_{\text{occ}}^{(\text{hole})}=n_{\bm{k}}^{2m-\alpha}|_{\text{unocc}}^{(\text{particle})}.
	\end{equation}
	Combining the particle and the hole contributions, we get
	\begin{align}
	&\kappa_{4}^{\text{BdG}} \nonumber\\
	&= \frac{1}{2\sqrt{2}}\sum_{\alpha=1,3,5,7}\sum_{\bm{k} \in K_4}e^{i\tfrac{\alpha\pi}{4}}(n_{\bm{k}}^{\alpha})^{\text{BdG}} \nonumber\\
	&= \frac{1}{2\sqrt{2}}\sum_{\alpha=1,3,5,7}e^{i\tfrac{\alpha\pi}{4}}\sum_{\bm{k} \in K_4}\left(n_{\bm{k}}^\alpha|_{\text{occ}}^{(\text{particle})}+n_{\bm{k}}^\alpha|_{\text{occ}}^{(\text{hole})}\right) \nonumber\\
	\label{SI_S4}
	&= \frac{1}{2\sqrt{2}}\sum_{\alpha=1,3,5,7}e^{i\tfrac{\alpha\pi}{4}}\sum_{\bm{k} \in K_4}\left(n_{\bm{k}}^\alpha|_{\text{occ}}^{(\text{particle})}+n_{\bm{k}}^{2m-\alpha}|_{\text{unocc}}^{(\text{particle})}\right) ,
	\end{align}
	
	where we denote the momenta invariant under $S_4$ by $K_4$ which are $(0,0,0), (\pi,\pi,0), (0,0,\pi),(\pi,\pi,\pi)$ in primitive lattices. As explained in the ``Inversion" section, the system is topologically trivial when all bands are occupied. In other words,
	\begin{align}
	\label{trivial_S4}
	 &\frac{1}{2\sqrt{2}}\sum_{\alpha=1,3,5,7}e^{i\tfrac{\alpha\pi}{4}}\sum_{\bm{k} \in K_4}\left(n_{\bm{k}}^\alpha|_{\text{occ}}^{(\text{particle})}+n_{\bm{k}}^\alpha|_{\text{unocc}}^{(\text{particle})} \right)\nonumber\\
	 &= 0\ \  (\text{mod}\ 2).
	\end{align}
	Then, the second term in Eq.\eqref{SI_S4} can be rewritten as
	\begin{widetext}
		\begin{align}
		\sum_{\alpha=1,3,5,7}e^{i\tfrac{\alpha\pi}{4}}\sum_{\bm{k} \in K_4}n_{\bm{k}}^{2m-\alpha}|_{\text{unocc}}^{(\text{particle})} &= e^{i\tfrac{m}{2}\pi}\sum_{x=1,3,5,7}e^{-i\tfrac{x\pi}{4}}\sum_{\bm{k} \in K_4}n_{\bm{k}}^{x}|_{\text{unocc}}^{(\text{particle})} \ \ (\text{mod}\ 2). 
		\end{align}
	Indeed, the relashionship $n^{\alpha}_{\bm{k}}|_{\text{occ}}^{(\text{particle})} = n^{-\alpha}_{\bm{k}}|_{\text{occ}}^{(\text{particle})}$ always holds when the TRS exists in the nomal phase. Substituting Eq.~\eqref{trivial_S4} and $n^{-x}_{\bm{k}}|_{\text{occ}}^{(\text{particle})} = n^{x}_{\bm{k}}|_{\text{occ}}^{(\text{particle})}$ for this equation,
		\begin{align}
		\sum_{\alpha=1,3,5,7}e^{i\tfrac{\alpha\pi}{4}}\sum_{\bm{k} \in K_4}n_{\bm{k}}^{2m-\alpha}|_{\text{unocc}}^{(\text{particle})} &= -e^{i\tfrac{m}{2}\pi}\sum_{x=1,3,5,7}e^{i\tfrac{x\pi}{4}}\sum_{\bm{k} \in K_4}n_{\bm{k}}^{x}|_{\text{unocc}}^{(\text{particle})} \ \ (\text{mod}\ 2). 
		\end{align}
	\end{widetext}
	
	If TRS also exists in superconducting phases, $\chi_{S_4}$ should be real (i.e. $m= 0,2$). As a result, $\kappa_{4}^{\text{BdG}}$ is always 0 (mod $2$) when $m=0$. On the other hand, when $m=2$, 
	\begin{align}
	\kappa_{4}^{\text{BdG}} &=\frac{1}{\sqrt{2}}\sum_{\alpha=1,3,5,7}\sum_{\bm{k} \in K_4}e^{i\tfrac{\alpha\pi}{4}}n_{\bm{k}}^{\alpha}|_{\text{occ}}^{(\text{particle})} \ \ (\text{mod}\ 2).
	\end{align}
	This reproduces Eq.(10) of the main text, where we omit labels ``(particle)" and ``occ."
	
	\section{The surface state for the $W=+2$ case}
	\label{App_B}
\begin{figure*}
		\begin{center}
			\includegraphics[width=2.0\columnwidth]{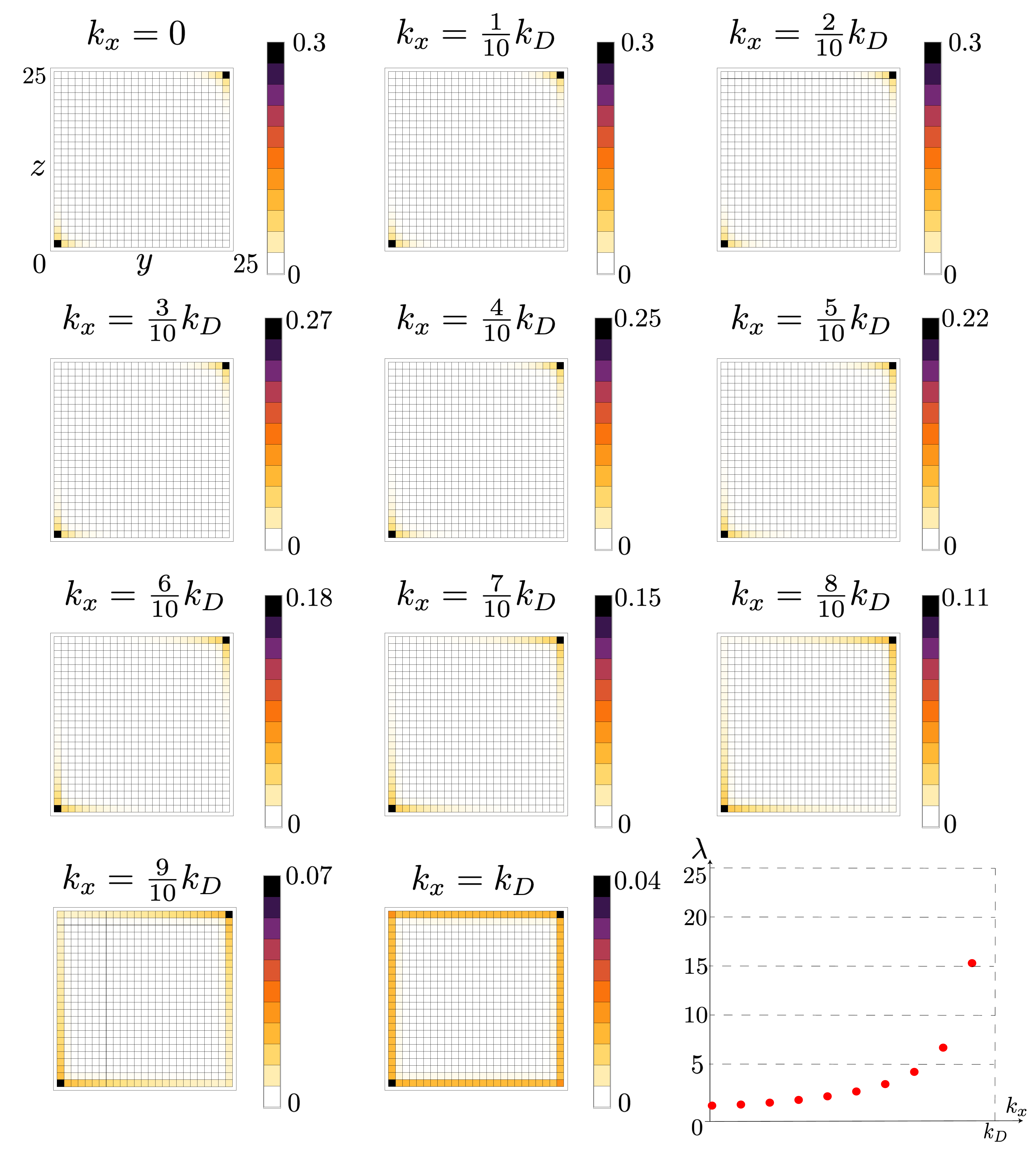}
			\caption{\label{LDOS} The average of real-space density profile of the four degenerate gapless states. The last panel represents the localization length of the gapless states along the edge. 
			}
		\end{center}
	\end{figure*}

	As discussed in the main text, the two copies of the ${}^3\text{He}$ model with the gap function $\Delta_{\bm{k}}^{(+,+)}$ shows the coexistence of the 1D and 2D surface states. Here plot their real-space density profile for several values of $k_x$.  Let $k_D>0$ be the momentum right at the Dirac point. (Numerically it was found to be $k_D=0.1151\pi$.) Figure~\ref{LDOS} plots the density profile for $k_x=\frac{i}{10}k_D$ $(i=0,1,\cdots,10)$. We see how the 1D hinge state evolves into 2D surface state.

	\clearpage
	
	\section{Detailed information of tight-binding models}
	\label{App_C}
	\subsection{$\text{Sr}_2\text{RuO}_4$}
	Here provide the three-orbit tight-binding model on a 2D square lattice~\cite{Sigrist} used in Fig.~1 in the main text. 
	\begin{widetext}
		\begin{eqnarray}
		&&H_{\bm{k}}=-
		\begin{pmatrix}
		(\mu+2t_1\cos k_y)\sigma_0 & 4t_4\sin k_x \sin k_y\sigma_0-i\lambda\sigma_z& i\lambda\sigma_y \\
		4t_4\sin k_x \sin k_y\sigma_0+i\lambda\sigma_z&(\mu+2t_1\cos k_x)\sigma_0 & -i\lambda\sigma_x \\
		-i\lambda\sigma_y  & i\lambda\sigma_x &[\mu'+2t_2(\cos k_x+\cos k_y)+4t_3\cos k_x \cos k_y]\sigma_0
		\end{pmatrix},\\
		&&\Delta_{\bm{k}}^{(1)}=\Delta^{(1)}(\sin k_x+i\sin k_y)
		\begin{pmatrix}
		\sigma_z i\sigma_y &0&0 \\
		0&\sigma_z i\sigma_y &0 \\
		0&0&\sigma_z i\sigma_y
		\end{pmatrix},\\
		&& \Delta_{\bm{k}}^{(2)}=\Delta^{(2)}
		\begin{pmatrix}
		(\sin k_x\sigma_x+\sin k_y\sigma_y)i\sigma_y &0&0 \\
		0&(\sin k_x\sigma_x+\sin k_y\sigma_y)i\sigma_y &0 \\
		0&0&(\sin k_x\sigma_x+\sin k_y\sigma_y)i\sigma_y
		\end{pmatrix}.
		\end{eqnarray}
	\end{widetext}
	The band structure is computed with the choice of parameters taken from Ref.~\onlinecite{PhysRevLett.111.087002}: $t_1=t_2=0.5$, $t_3=0.2$, $t_4=0.1$, $\mu = -0.2$, $\mu' = -0.2$, $\lambda =0.3$, $\Delta^{(1)}=0.6$, $\Delta^{(2)}=0$.  The four-fold rotation symmetry $C_{4}$, the mirror symmetry $M_{xy}$, and inversion symmetry $I$ are represented by
	\begin{eqnarray}
	U_{\bm{k}}(C_{4}) = \begin{pmatrix}  0 & e^{-i\tfrac{\pi}{4}\sigma_z} & 0 \\ -e^{-i\tfrac{\pi}{4}\sigma_z}& 0 & 0 \\ 0 & 0& -e^{-i\tfrac{\pi}{4}\sigma_z}\end{pmatrix},\\
	U_{\bm{k}}(M_{xy}) = \begin{pmatrix}  i\sigma_z & 0 & 0 \\ 0& i\sigma_z & 0 \\ 0 & 0& -i\sigma_z \end{pmatrix},\\
	U_{\bm{k}}(I) =\begin{pmatrix}  \sigma_0 & 0 & 0 \\ 0& \sigma_0 & 0 \\ 0 & 0& \sigma_0 \end{pmatrix},
	\end{eqnarray}
	which satisfy
	\begin{eqnarray}
	U_{\bm{k}}(C_{4})\Delta_{\bm{k}}^{(1)}U_{\bm{k}}(C_{4})^t = -i\Delta_{(-k_y,k_x,k_z)}^{(1)},\\
	U_{\bm{k}}(M_{xy}) \Delta_{\bm{k}}^{(1)}U_{\bm{k}}(M_{xy}) ^t = +\Delta_{(k_x, k_y, -k_z)}^{(1)},\\
	U_{\bm{k}}(I)\Delta_{\bm{k}}^{(1)}U_{\bm{k}}(I)^t = -\Delta_{-\bm{k}}^{(1)},\\
	U_{\bm{k}}(C_{4})\Delta_{\bm{k}}^{(2)}U_{\bm{k}}(C_{4})^t = +\Delta_{(-k_y,k_x,k_z)}^{(2)},\\
	U_{\bm{k}}(M_{xy}) \Delta_{\bm{k}}^{(2)}U_{\bm{k}}(M_{xy}) ^t = -\Delta_{(k_x, k_y, -k_z)}^{(2)},\\
	U_{\bm{k}}(I)\Delta_{\bm{k}}^{(2)}U_{\bm{k}}(I)^t = -\Delta_{-\bm{k}}^{(1)}.
	\end{eqnarray}
	
	In addition, the tight-binding model has two more mirror symmetries broken in the superconducting phase:
	\begin{eqnarray}
	U_{\bm{k}}(M_{yz}) = \begin{pmatrix}  -i\sigma_x & 0 & 0 \\ 0& i\sigma_x & 0 \\ 0 & 0& i\sigma_x \end{pmatrix},\\
	U_{\bm{k}}(M_{zx}) = \begin{pmatrix}  i\sigma_y& 0 & 0 \\ 0& -i\sigma_y & 0 \\ 0 & 0& i\sigma_y \end{pmatrix}.
	\end{eqnarray}
	
	
	In Table~\ref{TB_Sr2RuO4}, we summarize the rotation eigenvalues of the occupied bands in the normal phase.
	\begin{table*}
		\begin{center}
			\caption{\label{TB_Sr2RuO4}  The rotation eigenvalues and the number of occupied bands of $H_{\bm{k}}$ each high-symmetry point.
			}
			\begin{tabular}{c|cc}
				\hline\hline
				$\quad\bm{k}\quad$ & $\quad\quad\quad$$M_{xy}=+i$ sector$\quad\quad\quad$ & $\quad\quad\quad$$M_{xy}=-i$ sector$\quad\quad\quad$ \\
				\hline
				$\Gamma=(0,0)$ & $\xi^{+i}_{\Gamma}=e^{+\frac{3\pi i}{4}}$, $N^{+i}_{\Gamma}=3$ & $\xi^{-i}_{\Gamma}=e^{-\frac{3\pi i}{4}}$, $N^{-i}_{\Gamma}=3$\\
				$\mathrm{X}=(\pi,0)$ & $\zeta^{+i}_{\mathrm{X}}=i$, $N^{+i}_{\mathrm{X}}=1$ & $\zeta^{-i}_{\mathrm{X}}=-i$, $N^{-i}_{\mathrm{X}}=1$\\
				$\mathrm{M}=(\pi,\pi)$ & $\xi^{+i}_{\mathrm{M}}=1$, $N^{+i}_{\mathrm{M}}=0$ & $\xi^{-i}_{\mathrm{M}}=1$, $N^{-i}_{\mathrm{M}}=0$ \\
				\hline\hline
			\end{tabular}
		\end{center}
	\end{table*}
	
	For the $\chi_{C_4} = -i$ and $\chi_{M_{xy}} = +1$, the mirror chern numbers are 
	\begin{align}
	e^{\frac{2\pi i}{4}C^{\sigma = \pm i}} =(\chi_{C_4})^{2N_{X}^{-\sigma}- N_{\Gamma}^{-\sigma} - N_{M}^{-\sigma}}&\frac{\xi_{\Gamma}^{+\sigma}\xi_{M}^{+\sigma}}{\zeta_{X}^{+\sigma}}\frac{\xi_{\Gamma}^{-\sigma}\xi_{M}^{-\sigma}}{\zeta_{X}^{-\sigma}} \nonumber\\
	(C^{+i},C^{-i}) &=(1, 1) \mod 4.
	\end{align}

	\section{Transformation properties under spin rotation}
	Here we explain the transformation property of the perturbation term $V=-i\bm{m}\cdot\bm{\sigma}$ considered in the main text.
	
	Let us consider a rotation by an angle $\theta$ about an axis $\bm{n}$. The $\text{SO}(3)$ matrix representation of the rotation is given by $p=e^{-i\theta\bm{L}\cdot\bm{n}}$, where $\bm{L}=(L_1,L_2,L_3)^t$ is the matrix representation of the angular momentum and $(L_i)_{jk}=-i\epsilon_{ijk}$ ($\epsilon$ is the Levi-Civita tensor). The corresponding spin rotation is given by $p_{\text{sp}}=e^{-i\theta\bm{S}\cdot\bm{n}}$, where $\bm{S}=\frac{1}{2}\bm{\sigma}$ and $\bm{\sigma}$ is the Pauli matrix.
	
	Let us define $V_{\bm{m}}\equiv\bm{m}\cdot\bm{\sigma}$.  It satisfies $p_{\text{sp}}V_{\bm{m}}p_{\text{sp}}^{-1}=V_{p\bm{m}}$, meaning  that $\bm{m}$ transforms as a (pseudo-)vector.  Similarly, $\Delta_{\bm{d}}\equiv(\bm{d}\cdot\bm{\sigma})i\sigma_y$ satisfies $p_{\text{sp}}\Delta_{\bm{d}}p_{\text{sp}}^{t}=\Delta_{p\bm{d}}$.  Thus $\bm{d}$ in $\Delta_{\bm{d}}$ also transforms as a (pseudo-)vector (but remember $p_{\text{sp}}^{-1}$ is replaced by $p_{\text{sp}}^t$) .
	
	\section{How to transform irreducible representations in body-centered lattices to irreducible representations in primitive lattices}
	\label{App_E}
	In the main text, we discussed the indicators for space group $I4/m$, a body-centered system, using the Brillouin zone of the primitive system $P4/m$, which has a simpler structure. Here we summarize the conversion rule. Here, we summarize the conversion rule (Table~\ref{conversion}).
	
	Let us denote the Brillouin zone for $I4/m$ and $P4/m$ by $B_I$ and $B_P$, respectively. We assume $(\hat{C}_n)^n=(\hat{M}_{xy})^2=\eta^{N_F}$, where $\hat{C}_n$ is the $n$-fold rotation about $z$-axis, $\hat{M}_{xy}$ is the mirror about the $xy$ plane, $\eta=\pm1$ distinguishes the spinless fermions ($+1$) and spinful fermions ($-1$), and $N_F=+1$ for single-particle problems.  We also assume that the rotation $\hat{C}_{n}$, the mirror $\hat{M}_{xy}$, and the inversion symmetry $\hat{I}$ all commute and that $(\hat{C}_{4})^2=\hat{C}_{2}$.
	
	In the following, we denote the $C_{4}$ eigenvalue by $\xi$, the $C_{2}$ eigenvalue by $\zeta$, and the $M_{xy}$ eigenvalue by $\sigma$.
	
	\begin{table*}
		\begin{center}
			\begin{tabular}{c|c|c|c}
				\hline\hline
				Coordinate in $B_I$ & irreducible representations of $I4/m$ & Coordinate in $B_P$ & irreducible representations of $P4/m$\\
				\hline
				$\Gamma$: $(0, 0, 0)$ & $(C_{4}, M_{xy}) = (\xi_{\Gamma}, \sigma_{\Gamma})$&  \multirow{2}{*}{$(0, 0, 0)$} &  \multirow{2}{*}{$(C_{4}, M_{xy}) = (\xi_{\Gamma}, \sigma_{\Gamma}), (\xi_{\mathrm{Z}}, \sigma_{\mathrm{Z}})$}\\
				$\mathrm{Z}$: $(0, 0, 2\pi)$ & $(C_{4}, M_{xy}) = (\xi_{\mathrm{Z}}, \sigma_{\mathrm{Z}})$ & & \\
				\hline
				$(\pi, 0, 0)$ & $M_{xy} = \sigma$ & $(\pi, 0, 0)$ & $(C_{2}, M_{xy}) = (\sqrt{\eta}, \sigma),  (-\sqrt{\eta}, \sigma)$\\
				\hline
				$\mathrm{X}$: $(\pi, \pi, 0)$ & $(C_{2}, M_{xy}) = (\zeta, \sigma)$ & $(\pi, \pi, 0)$ & $(C_{4}, M_{xy}) = (\sqrt{\zeta}, \sigma),  (-\sqrt{\zeta}, \sigma)$\\
				\hline
				$(0, 0, \pi)$ & $C_{4} = \xi$ & $(0, 0, \pi)$ &$(C_{4}, M_{xy}) = (\xi, \sqrt{\eta}), (\xi, -\sqrt{\eta})$ \\
				\hline
				$\mathrm{N}$: $(\pi, 0, \pi)$ & $I = p$ & $(\pi, 0, \pi)$ & $(C_{2}, M_{xy}) = (\sqrt{\eta}, p\sqrt{\eta}),  (-\sqrt{\eta}, -p\sqrt{\eta})$\\
				\hline
				$\mathrm{P}$: $(\pi, \pi, \pi)$ & $(S_{4})^{3} = q$& $(\pi, \pi, \pi)$ & $(C_{4}, M_{xy}) = (\sqrt{\eta q^2}, q^{-1}\eta\sqrt{\eta q^2}), (-\sqrt{\eta q^2}, -q^{-1}\eta\sqrt{\eta q^2}) $ \\
				\hline\hline
			\end{tabular}
			\caption{\label{conversion}The conversion table of irres between $I4/m$ and $P4/m$.}
		\end{center}
	\end{table*}

	\subsection{$(0,0,0)$}
	The PG of $\Gamma, \mathrm{Z}\in B_I$ is $D_{4h}$. 
	Therefore, states at $\Gamma, \mathrm{Z}\in B_I$ can be labeled by $\xi$ and $\sigma$. The two momenta in $B_I$ merge to a single momentum $(0,0,0)\in B_P$. Hence,
	\begin{quote}
		\emph{The two 1D representations $(C_{4}=\xi_{\Gamma}, M_{xy}=\sigma_{\Gamma})$ at $\Gamma \in B_I$ and $(C_{4}=\xi_{\mathrm{Z}}, M_{xy}=\sigma_{\mathrm{Z}})$ at $\mathrm{Z} \in B_I$ reduces to two 1D representations $(C_{4}=\xi_{\Gamma},  
			M_{xy}=\sigma_{\Gamma})$ and $(C_{4}=\xi_{\mathrm{Z}},  M_{xy}=\sigma_{\mathrm{Z}})$ at $(0,0,0)\in B_P$ for $P4/m$.}
	\end{quote}
	
	\subsection{$(\pi, 0, 0)$}
	The PG of $(\pi, 0, 0)\in B_I$ is $C_s$ and states at $(\pi,0,0)\in B_I$ can be labeled by $\sigma$. We denote them by $|\sigma\rangle_{(\pi,0,0)}$:
	\begin{eqnarray}
	\hat{M}_{xy}|\sigma\rangle_{(\pi,0,0)}=\sigma |\sigma\rangle_{(\pi,0,0)}.
	\end{eqnarray}
	Note that $C_{2}$ maps $(\pi,0,0)$ to $(-\pi,0,0)$. These two points are different in $B_I$ but are identical in $B_P$. Thus we have
	\begin{eqnarray}
	\hat{C}_{2}|\sigma\rangle_{(\pi,0,0)}\equiv |\sigma\rangle_{(-\pi,0,0)},\\
	\hat{C}_{2}|\sigma\rangle_{(-\pi,0,0)}=(\hat{C}_{2})^2|\sigma\rangle_{(\pi,0,0)}=\eta|\sigma\rangle_{(\pi,0,0)}.
	\end{eqnarray}
	Therefore, $C_{2}$ is represented by $U_{(\pi,0,0)}(C_{2})=\begin{pmatrix}0&1\\\eta&0\end{pmatrix}$, whose eigenvalues are $\pm \sqrt{\eta}$. This means that
	\begin{quote}
		\emph{The 1D representation $M_{xy}=\sigma$ at $(\pi,0,0)\in B_I$ reduces to two 1D representations $(C_{2}=\sqrt{\eta}, M_{xy}=\sigma)$ and $(C_{2}=-\sqrt{\eta}, M_{xy}=\sigma)$ at $(\pi,0,0)\in B_P$.}
	\end{quote}
	
	\subsection{$(\pi,\pi,0)$}
	The PG of $\mathrm{X}\in B_I$ is $D_{2h}$. Therefore, states at $\mathrm{X}\in B_I$ can be labeled by $\zeta$ and $\sigma$. We denote them by $|\zeta,\sigma\rangle_{\mathrm{X}}$:
	\begin{eqnarray}
	\hat{C}_{2}|\zeta,\sigma\rangle_{\mathrm{X}}=\zeta |\zeta,\sigma\rangle_{\mathrm{X}},\quad
	\hat{M}_{xy}|\zeta,\sigma\rangle_{\mathrm{X}}=m |\zeta,\sigma\rangle_{\mathrm{X}}.
	\end{eqnarray}
	Note that $C_{4}$ maps $\mathrm{X}=(\pi,\pi,0)$ to $\mathrm{X}'=(-\pi,\pi,0)$, which are distinct in $B_I$ but are identical in $B_P$:
	\begin{eqnarray}
	\hat{C}_{4}|\zeta,\sigma\rangle_{\mathrm{X}}\equiv |\zeta,\sigma\rangle_{\mathrm{X}'},\\
	\hat{C}_{4}|\zeta,\sigma\rangle_{\mathrm{X}'}=(\hat{C}_{4})^2|\zeta,\sigma\rangle_{\mathrm{X}}=\zeta|\zeta,\sigma\rangle_{\mathrm{X}}.
	\end{eqnarray}
	Therefore, $C_{4}$ is represented by $U_{(\pi,\pi,0)}(C_{4})=\begin{pmatrix}0&1\\\zeta&0\end{pmatrix}$, whose eigenvalues are $\pm \sqrt{\zeta}$. This means that
	\begin{quote}
		\emph{The 1D representation $(C_{2}=\zeta, M_{xy}=\sigma)$ at $\mathrm{X}\in B_I$ reduces to two 1D representations $(C_{4}=\sqrt{\zeta}, M_{xy}=\sigma)$ and $(C_{4}=-\sqrt{\zeta}, M_{xy}=\sigma)$ at $(\pi,\pi,0)\in B_P$.}
	\end{quote}
	
	\subsection{$(0,0,\pi)$}
	The PG of $(0, 0, \pi)\in B_I$ is $C_{4v}$. Therefore, states at $(0,0,\pi)\in B_I$ can be labeled by $\xi$. We denote them by $|\xi\rangle_{(0,0,\pi)}$:
	\begin{eqnarray}
	\hat{C}_{4}|\xi\rangle_{(0,0,\pi)}=\xi |\xi\rangle_{(0,0,\pi)}.
	\end{eqnarray}
	Note that $M_{xy}$ maps $(0,0,\pi)$ to $(0,0,-\pi)$:
	\begin{eqnarray}
	\hat{M}_{xy}|\xi\rangle_{(0,0,\pi)}\equiv |\xi\rangle_{(0,0,-\pi)}, \\
	\hat{M}_{xy}|\xi\rangle_{(0,0,-\pi)}=(\hat{C}_{2})^2|\xi\rangle_{(0,0,\pi)}=\eta|\xi\rangle_{(0,0,\pi)}.
	\end{eqnarray}
	Therefore, $M_{xy}$ is represented by $U_{(0,0,\pi)}(M_{xy})=\begin{pmatrix}0&1\\\eta&0\end{pmatrix}$
	, whose eigenvalues are $\pm \sqrt{\eta}$. This means that
	\begin{quote}
		\emph{The 1D representation $C_4=\xi$ at $(0,0,\pi)\in B_I$ reduces to two 1D representations $(C_{4}=\xi, M_{xy}=\sqrt{\eta})$ and $(C_{4}=\xi, M_{xy}=-\sqrt{\eta})$ at $(0,0,\pi)\in B_P$.}
	\end{quote}
	
	\subsection{$(\pi,0,\pi)$}
	The PG of $\mathrm{N}\in B_I$ is $C_{2h}$. Therefore, states at $\mathrm{N}\in B_I$ can be labeled by the parity eigenvalue $p=\pm1$.  $\hat{M}_{xy}\hat{C}_{2}=\eta^{N_F} \hat{I}$. 
	\begin{eqnarray}
	\hat{M}_{xy}\hat{C}_{2}|p\rangle_{\mathrm{N}}=\eta p |p\rangle_{\mathrm{N}}.
	\end{eqnarray}
	$C_{2}$ and $M_{xy}$ map $\mathrm{N}=(\pi,0,\pi)$ to $\mathrm{N}'=(-\pi,0,\pi)=(\pi,0,-\pi)$:
	\begin{align}
	\hat{C}_{2}|p\rangle_{\mathrm{N}}&\equiv |p\rangle_{\mathrm{N'}},\\ \hat{C}_{2}|p\rangle_{\mathrm{N'}}&=(\hat{C}_{2})^2|p\rangle_{\mathrm{N}}=\eta|p\rangle_{\mathrm{N}},\\
	\hat{M}_{xy}|p\rangle_{\mathrm{N'}}&=\hat{M}_{xy}\hat{C}_{2}|p\rangle_{\mathrm{N}}=\eta p|p\rangle_{\mathrm{N}},\\ \hat{M}_{xy}|p\rangle_{\mathrm{N}}&=\eta p(\hat{M}_{xy})^2|p\rangle_{\mathrm{N'}}=p|p\rangle_{\mathrm{N'}},
	\end{align}
	Therefore, $C_{2}$ and $M_{xy}$ are represented by $U_{(\pi,0,\pi)}(C_{2})=\begin{pmatrix}0&1\\\eta&0\end{pmatrix}$ and $U_{(\pi,0,\pi)}(M_{xy})=p U_{(\pi,0,\pi)}(C_{2})$. Therefore,
	\begin{quote}
		\emph{The 1D representation $\hat{M}_{xy}\hat{C}_{2}=p$ at $\mathrm{N}\in B_I$ reduces to two 1D representations $(C_{2}=\sqrt{\eta}, M_{xy}=p\sqrt{\eta})$ and $(C_{2}=-\sqrt{\eta}, M_{xy}=-p\sqrt{\eta})$ at $(\pi,0,\pi)\in B_{P}$.}
	\end{quote}
	
	\subsection{$(\pi,\pi,\pi)$}
	The PG of $\mathrm{P}\in B_I$ is $D_{2d}$. Therefore, states at $\mathrm{P}\in B_I$ can be labeled by the eigenvalue $q$ of the product $M_{xy}C_{4}$:
	\begin{eqnarray}
	\hat{M}_{xy}\hat{C}_{4}|q\rangle_{\mathrm{P}}=q |q\rangle_{\mathrm{P}}.
	\end{eqnarray}
	$C_{4}$ and $M_{xy}$ map $\mathrm{P}=(\pi,\pi,\pi)$ to $\mathrm{P}'=(-\pi,\pi,\pi)=(\pi,\pi,-\pi)$:
	\begin{align}
	\hat{C}_{4}|q\rangle_{\mathrm{P}}&\equiv |q\rangle_{\mathrm{P}'},\\ 
	\hat{C}_{4}|q\rangle_{\mathrm{P}'}&=(\hat{C}_{4})^2|q\rangle_{\mathrm{P}}=\eta q^2|q\rangle_{\mathrm{P}},\\
	\hat{M}_{xy}|q\rangle_{\mathrm{P}'}&=\hat{M}_{xy}\hat{C}_{4}|q\rangle_{\mathrm{P}}=q|q\rangle_{\mathrm{P}},\\
	\hat{M}_{xy}|q\rangle_{\mathrm{P}}&=q^{-1}(\hat{M}_{xy})^2|p\rangle_{\mathrm{P}'}=q^{-1}\eta|p\rangle_{\mathrm{P}'}.
	\end{align}
	Therefore, $C_{4}$ and $M_{xy}$ are represented by $U_{(\pi,\pi,\pi)}(C_{4})=\begin{pmatrix}0&1\\\eta q^2&0\end{pmatrix}$ and $U_{(\pi,\pi,\pi)}(M_{xy})=q^{-1}\eta U_{(\pi,\pi,\pi)}(C_{4})$.
	\begin{quote}
		\emph{The 1D representation $\hat{M}_{xy}\hat{C}_{4}=q$ at $\mathrm{P}\in B_I$ reduces to two 1D representations $(C_{4}=\sqrt{\eta q^2}, M_{xy}=q^{-1}\eta \sqrt{\eta q^2})$ and $(C_{4}=-\sqrt{\eta q^2}, M_{xy}=-q^{-1}\eta \sqrt{\eta q^2})$ at $(\pi,\pi,\pi)\in B_{P}$.}
	\end{quote}

	\section{DFT results}
	\label{App_F}
	Here we present the details on our DFT calculations. We include the spin-orbital coupling and adopt the standard generalized gradient approximation (GGA) with Perdew-Burke-Ernzerhof (PBE) realization for the exchange-correlation functional~\cite{PhysRevLett.77.3865}. 
	
	\subsection{Crystal structures and band structures}
\begin{figure}[H]
		\begin{center}
			\begin{tabular}{c}			
				\begin{minipage}{0.49\hsize}
					\begin{center}	
						\includegraphics[width=0.9\columnwidth]{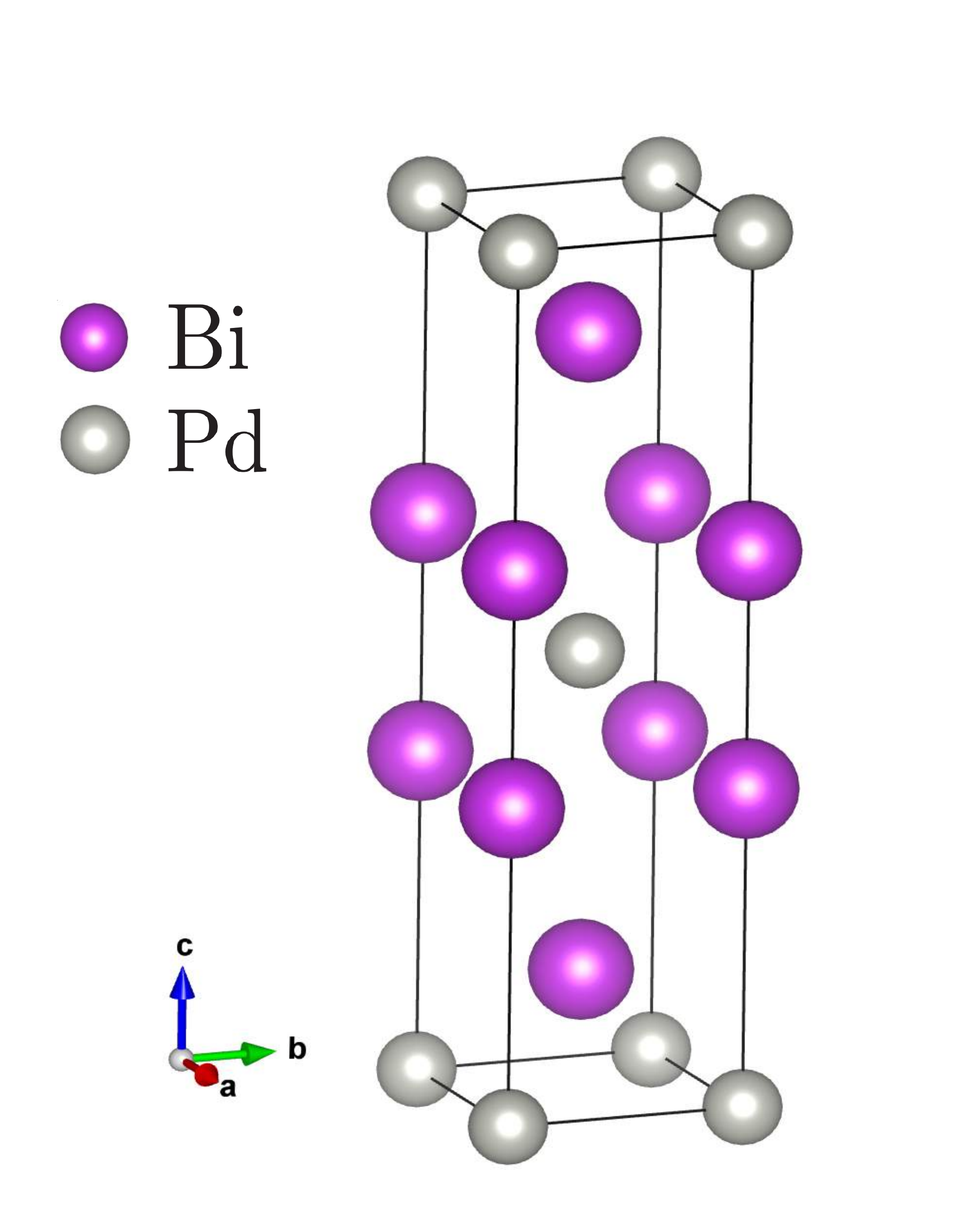}
						\caption{\textbf{Crystal structure of $\beta\text{-PdBi}_2$ \\ (mp-570197).}\label{CS_b-PdBi2}}
					\end{center}
				\end{minipage}
				\begin{minipage}{0.49\hsize}
					\begin{center}	
						\includegraphics[width=0.8\columnwidth]{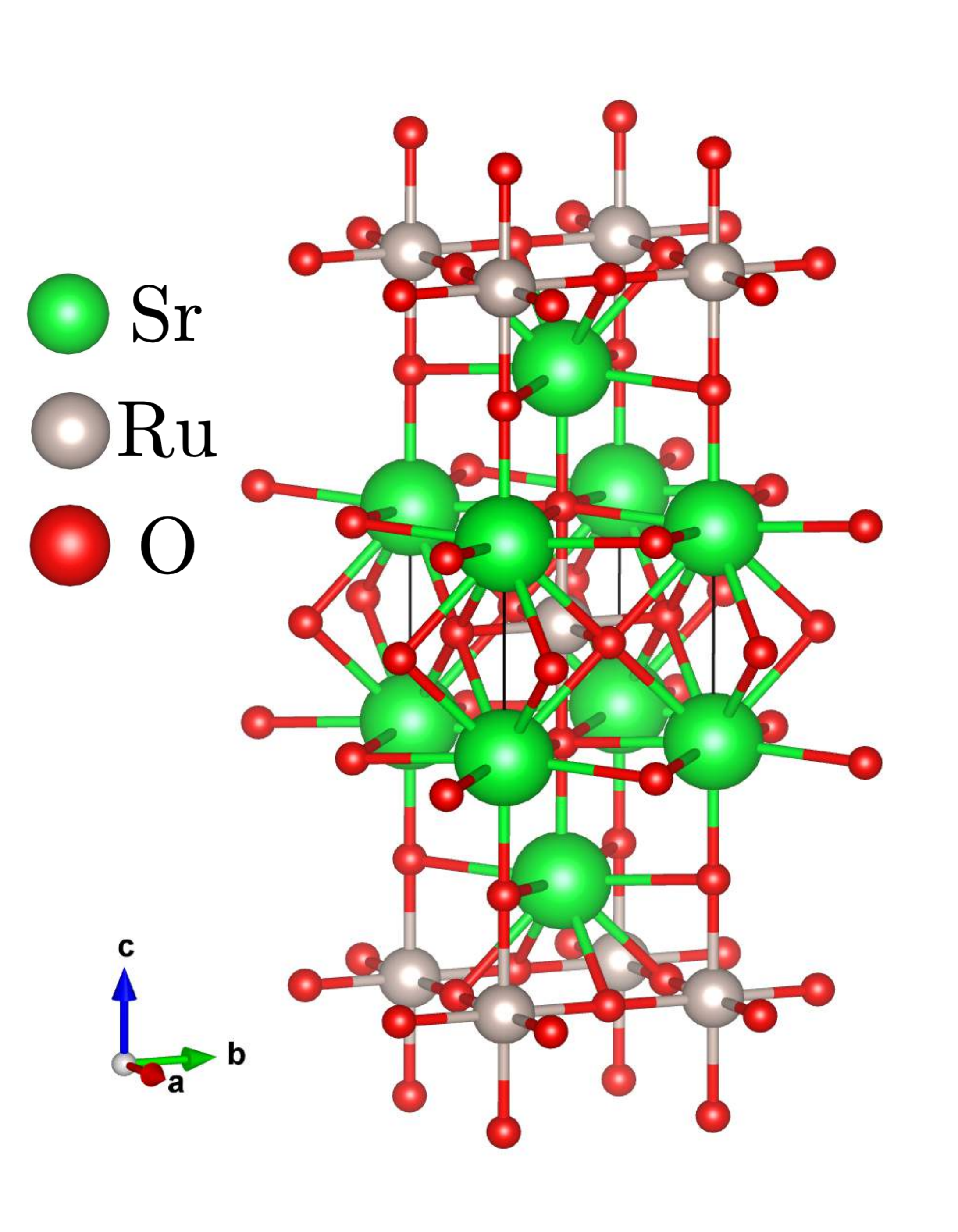}
						\caption{\textbf{Crystal structure of $\text{Sr}_2\text{Ru}\text{O}_4$\\ (mp-4596).}\label{CS_Sr2RuO4}}
					\end{center}
				\end{minipage}
			\end{tabular}
		\end{center}
	\end{figure}

\begin{figure*}
		\begin{center}
			\begin{tabular}{c}			
				\begin{minipage}{0.495\hsize}
					\begin{center}	
						\includegraphics[width=0.8\columnwidth]{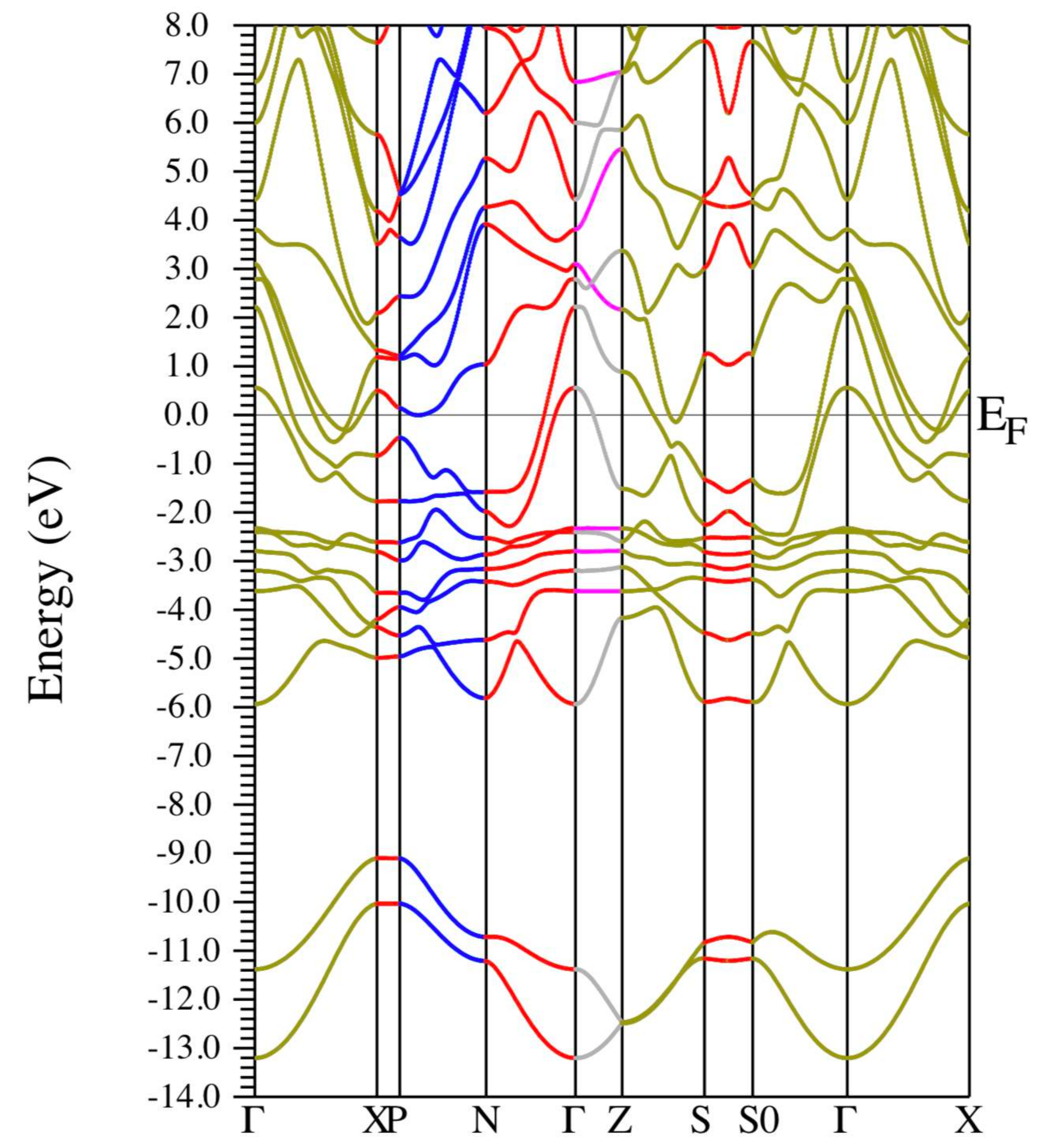}
						\caption{\textbf{Band structure of $\beta\text{-PdBi}_2$.}\label{BS_b-PdBi2}}
					\end{center}
				\end{minipage}
				\begin{minipage}{0.495\hsize}
					\begin{center}
						\includegraphics[width=0.8\columnwidth]{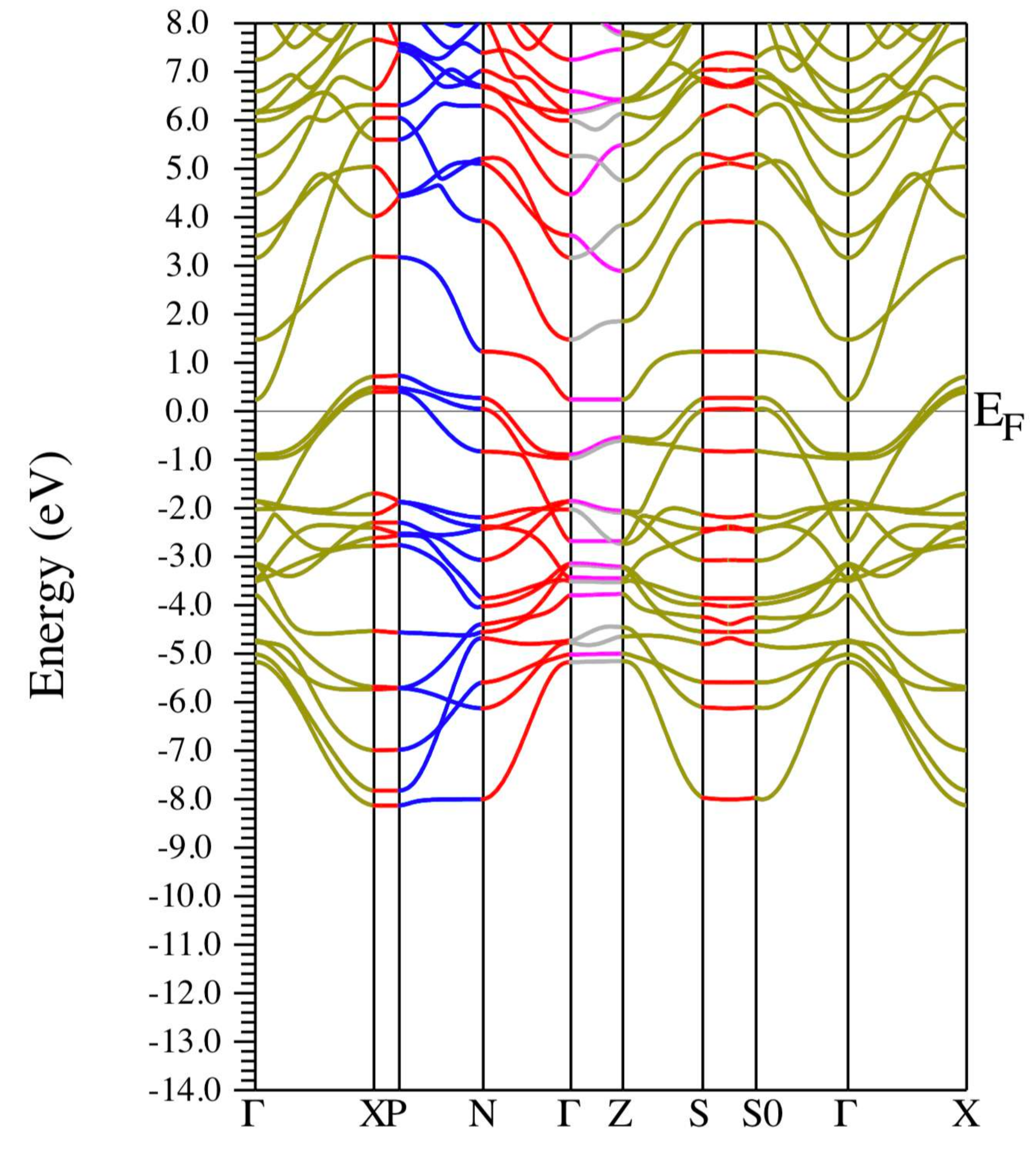}
						\caption{\textbf{Band structure of $\text{Sr}_2\text{Ru}\text{O}_4$.}\label{BS_Sr2RuO4}}
					\end{center}
				\end{minipage}
			\end{tabular}
		\end{center}
	\end{figure*}

	The crystal structure of each material is illustrated in Figs.~\ref{CS_b-PdBi2}-\ref{CS_Sr2RuO4}.
	The Materials Project ID~\cite{Jain2013} is also indicated in the caption. The corresponding band structure is given in Figs.~\ref{BS_b-PdBi2}--\ref{BS_Sr2RuO4}.
	
	\subsection{The list of irreducible representations and the calculation precess of symmetry indicators}
	Here we summarize the irreducible representations (irrep) of the point group (PG) $\mathcal{G}_{\bm{k}}/T$ at high-symmetry points for $I4/mmm$ and $I4/m$. In Tables~\ref{irrep_b-PdBi2}--\ref{irrep_Sr2RuO4_SC}, ``$\oplus$" between two irreducible representations indicates that the two irreducible representations are paired under the TRS. 
	We also list $n_{\bm{k}}^{\alpha}$ at high-symmetry momenta $\bm{k}$ for each material.  
	\begin{table*}
		\begin{center}
			\caption{\label{irrep_b-PdBi2} The list of irreducible representations of the PG at high-symmetry points for $I4/mmm$. The last row shows $n_{k}^{\alpha}$ for $\beta\text{-PdBi}_2$.}
			\begin{tabular}{c|cccc|cccc|cc|cc|cc}
				\hline\hline
				$\bm{k}$ & \multicolumn{4}{c|}{$\Gamma$: $(0, 0, 0)$}    &  \multicolumn{4}{c|}{$\mathrm{Z}$: $(0, 0, 2\pi)$}   & \multicolumn{2}{c|}{X: $(\pi, \pi, 0)$}  &  \multicolumn{2}{c|}{N: $(\pi, 0, \pi)$} & \multicolumn{2}{c}{$\mathrm{P}$: $(\pi, \pi, \pi)$}\\
				\hline
				PG. & \multicolumn{4}{c|}{$D_{4h}$}    &  \multicolumn{4}{c|}{$D_{4h}$}   & \multicolumn{2}{c|}{$D_{2h}$}  &  \multicolumn{2}{c|}{$C_{2h}$} & \multicolumn{2}{c}{$D_{2d}$} \\
				\hline
				Irrep &$\Gamma_{6}^{+}$ & $\Gamma_{7}^{+}$ & $\Gamma_{6}^{-}$&$\Gamma_{7}^{-}$&$\Gamma_{6}^{+}$ & $\Gamma_{7}^{+}$ & $\Gamma_{6}^{-}$&$\Gamma_{7}^{-}$&$\Gamma_{5}^{+}$  &$\Gamma_{5}^{-}$ & $\Gamma_{3}^{+} \oplus \Gamma_{4}^{+}$  & $\Gamma_{3}^{-} \oplus \Gamma_{4}^{-}$  & $\Gamma_{6}$ & $\Gamma_{7}$\\
				\hline
				$n_{\bm{k}}^{\alpha}$ &$6$&$6$&$5$&$4$&$6$&$6$&$6$&$4$&$12$&$11$&$12$&$11$&$10$&$13$\\
				\hline\hline
			\end{tabular}
		\end{center}
	\end{table*}

	\begin{table*}
		\begin{center}
			\caption{\label{irrep_Sr2RuO4} 
				The same as Table~\ref{irrep_b-PdBi2} for space group $I4/mmm$ and for material  $\text{Sr}_2\text{Ru}\text{O}_4$.}
			\begin{tabular}{c|cccc|cccc|cc|cc|cc|cc|c}
				\hline\hline
				$\bm{k}$ & \multicolumn{4}{c|}{$\Gamma$: $(0, 0, 0) $}    &  \multicolumn{4}{c|}{$\mathrm{Z}$: $(0, 0, 2\pi)$}   & \multicolumn{2}{c|}{$\mathrm{X}$: $(\pi, \pi, 0)$}  &  \multicolumn{2}{c|}{$\mathrm{N}$: $(\pi, 0, \pi)$} & \multicolumn{2}{c|}{$\mathrm{P}$: $(\pi, \pi, \pi)$} & \multicolumn{2}{c|}{$(0, 0, \pi)$} & $(\pi, 0, 0)$ \\
				\hline
				PG & \multicolumn{4}{c|}{$D_{4h}$}    &  \multicolumn{4}{c|}{$D_{4h}$}   & \multicolumn{2}{c|}{$D_{2h}$}  &  \multicolumn{2}{c|}{$C_{2h}$} & \multicolumn{2}{c|}{$D_{2d}$} & \multicolumn{2}{c|}{$C_{4v}$} & $C_s$\\
				\hline
				Irrep &$\Gamma_{6}^{+}$ & $\Gamma_{7}^{+}$ & $\Gamma_{6}^{-}$&$\Gamma_{7}^{-}$&$\Gamma_{6}^{+}$ & $\Gamma_{7}^{+}$ & $\Gamma_{6}^{-}$&$\Gamma_{7}^{-}$&$\Gamma_{5}^{+}$  &$\Gamma_{5}^{-}$ & $\Gamma_{3}^{+} \oplus \Gamma_{4}^{+}$  & $\Gamma_{3}^{-} \oplus \Gamma_{4}^{-}$  & $\Gamma_{6}$ & $\Gamma_{7}$ & $\Gamma_{6}$ & $\Gamma_{7}$ & $\Gamma_5$\\
				\hline
				$n_{\bm{k}}^{\alpha}$ &$9$&$5$&$11$&$6$&$9$&$5$&$11$&$6$&$15$&$13$&$14$&$15$&$14$&$14$&$20$&$11$&$31$\\
				\hline\hline
			\end{tabular}
		\end{center}
	\end{table*}
	DFT gives us irreducible representations in the nomal-conducting phase. However, as explained in the main text, the time-reversal symmetry and two mirror symmetries are broken in the SC phase. In other words, the space group in the SC phase is $I4/m$. Therefore, we should calculate irreducible representations in SC phase by group theory called compatibility relations. Here, $H$ is subgroup of $G$, and $D_{a}^{H}$ and $D_{b}^{G}$ are an irrep of $H$ and an irrep of $G$, respevtively.  Suppose that $N_H$, $\chi(h)$ are the number of components of $H$ and characters, respectively. Then, we can know how many $D_{a}^{H}$ are included in $D_{b}^{G}$ from 
	\begin{eqnarray}
	\frac{1}{N_H}\sum_{h \in H} \chi^{*}_{D_{a}^{H}}(h)\chi_{D_{b}^{G}}(h)
	\end{eqnarray}
	When PG $D_{4h}$ changes to $C_{4h}$ at $\Gamma$ and $Z$, irreducible representations are decomposed as below. 
	\begin{eqnarray}
	\Gamma_{6}^{+} \rightarrow \Gamma_{5}^{+} \oplus \Gamma_{6}^{+} \\
	\Gamma_{7}^{+} \rightarrow \Gamma_{7}^{+} \oplus \Gamma_{8}^{+}\\
	\Gamma_{6}^{-} \rightarrow \Gamma_{5}^{-} \oplus \Gamma_{6}^{-} \\
	\Gamma_{7}^{-} \rightarrow  \Gamma_{7}^{-} \oplus \Gamma_{8}^{-} 
	\end{eqnarray}
	When PG $D_{2h}$ changes to $C_{2h}$ at $X$, irreducible representations are decomposed as below. 
	\begin{eqnarray}
	\Gamma_{5}^{+} \rightarrow \Gamma_{3}^{+} \oplus \Gamma_{4}^{+} \\
	\Gamma_{5}^{-} \rightarrow \Gamma_{3}^{-} \oplus \Gamma_{4}^{-} 
	\end{eqnarray}
	
	When PG $D_{2d}$ changes to $S_{4}$ at $P$, irreducible representations are decomposed as below.
	\begin{eqnarray}
	\Gamma_6 \rightarrow \Gamma_5 \oplus \Gamma_6 \\
	\Gamma_7 \rightarrow \Gamma_7 \oplus \Gamma_8
	\end{eqnarray}
	
	When PG $C_{2h}$ changes to $C_i$ at $N$, irreducible representations are decomposed as below.
	\begin{eqnarray}
	\Gamma_{3}^{+} \rightarrow \Gamma_{2}^{+} \\
	\Gamma_{4}^{+} \rightarrow \Gamma_{2}^{+} \\
	\Gamma_{3}^{-} \rightarrow \Gamma_{2}^{-} \\
	\Gamma_{4}^{-} \rightarrow \Gamma_{2}^{-}
	\end{eqnarray}
	
	When PG $C_{2v}$ changes to $C_s$ at $(\pi,0,0)$, irreducible representations are decomposed as below
	\begin{eqnarray}
	\Gamma_5 \rightarrow \Gamma_3 \oplus \Gamma_4
	\end{eqnarray}.
	
	When PG $C_{4v}$ changes to $C_{4}$ at $(0,0,\pi)$, irreducible representations are decomposed as below. 
	\begin{eqnarray}
	\Gamma_6 \rightarrow \Gamma_5 \oplus \Gamma_6 \\
	\Gamma_7 \rightarrow \Gamma_7 \oplus \Gamma_8
	\end{eqnarray}
	
	\begin{table*}
		\begin{center}
			\caption{\label{irrep_Sr2RuO4_SC} 
				irreducible representations for space group $I4/m$ and for material  $\text{Sr}_2\text{Ru}\text{O}_4$ in the SC phase.}
			\begin{tabular}{c|cccc|cc|cc|cc|cc|c}
				\hline\hline
				$\bm{k}$ & \multicolumn{4}{c|}{$\Gamma$: $(0, 0, 0)$, $\mathrm{Z}$: $(0, 0, 2\pi)$}   & \multicolumn{2}{c|}{$\mathrm{X}$: $(\pi, \pi, 0)$}  &  \multicolumn{2}{c|}{$\mathrm{N}$: $(\pi, 0, \pi)$} & \multicolumn{2}{c|}{$\mathrm{P}$: $(\pi, \pi, \pi)$} & \multicolumn{2}{c|}{$(0, 0, \pi)$} & $(\pi, 0, 0)$ \\
				\hline
				PG & \multicolumn{4}{c|}{$C_{4h}$} & \multicolumn{2}{c|}{$C_{2h}$}  &  \multicolumn{2}{c|}{$C_{i}$} & \multicolumn{2}{c|}{$S_{4}$} & \multicolumn{2}{c|}{$C_{4}$} & $C_s$\\
				\hline
				Irrep &$\Gamma_{5}^{+}\oplus\Gamma_{6}^{+}$ & $\Gamma_{7}^{+}\oplus\Gamma_{8}^{+}$ & $\Gamma_{5}^{-}\oplus\Gamma_{6}^{-}$&$\Gamma_{7}^{-}\oplus\Gamma_{8}^{-}$&$\Gamma_{3}^{+}\oplus\Gamma_{4}^{+}$ & $\Gamma_{3}^{-}\oplus\Gamma_{4}^{-}$ & $\Gamma_{2}^{+}$  & $\Gamma_{2}^{-}$ & $\Gamma_{5}\oplus\Gamma_{6}$ & $\Gamma_{7}\oplus\Gamma_{8}$ & $\Gamma_{5}\oplus\Gamma_{6}$ & $\Gamma_{7}\oplus\Gamma_{8}$ & $\Gamma_{3}\oplus\Gamma_{4}$\\
				\hline
				$n_{\bm{k}}^{\alpha}$ &$9$&$5$&$11$&$6$&$15$&$13$&$28$&$30$&$14$&$14$&$20$&$11$&$31$\\
				\hline\hline
			\end{tabular}
		\end{center}
	\end{table*}
	
	From the results summarized in Table IV, the irreducible representations of $I4/m$ are transformed into those of $P4/m$. We list the numbers of irreducible representations after the transformation in Table~\ref{tab:number_gamma_Sr2RuO4_83} and~\ref{tab:number_x_Sr2RuO4_3}.
	\begin{table}[H]
		\begin{center}
				\caption{the number of irreducible representations at $\Gamma:(0,0,0)$, $M:(\pi, \pi, \pi)$, $Z:(0,0,\pi)$, and $A:(\pi,\pi,\pi)$ in $P4/m$ (PG $C_{4h}$)
				\label{tab:number_gamma_Sr2RuO4_83}}
					\begin{tabular}{c||c|c|c|c|c|c|c|c}
					\hline \hline
					k-points & $n_{\mathrm{occ}}^{5_{+},p}$ & $n_{\mathrm{occ}}^{6_{+},p}$ & $n_{\mathrm{occ}}^{7_{+},p}$ & $n_{\mathrm{occ}}^{8_{+},p}$ & $n_{\mathrm{occ}}^{5_{-},p}$ & $n_{\mathrm{occ}}^{6_{-},p}$ & $n_{\mathrm{occ}}^{7_{-},p}$ & $n_{\mathrm{occ}}^{8_{-},p}$\\
					\hline
					$\Gamma$ & $18$ & $18$& $10$ & $10$ & $22$& $22$ & $12$ & $12$\\
					$M$ & $15$ & $15$& $15$ & $15$ & $13$& $13$ & $13$ & $13$\\
					$Z$ & $20$ & $20$& $11$ & $11$ & $20$& $20$ & $11$ & $11$\\
					$A$ & $14$ & $14$& $14$ & $14$ & $14$& $14$ & $14$ & $14$\\
					\hline\hline
				\end{tabular}
			\end{center}
	\end{table}

	\begin{table}[H]
		\begin{center}
		\caption{the number of irreducible representations at $X:(\pi,0,0)$ and $R:(\pi,0,\pi)$ in $P4/m$ (PG $C_{2h}$)
			\label{tab:number_x_Sr2RuO4_3}}
		\begin{tabular}{c||c|c|c|c}
			\hline \hline
			k-points & $n_{\mathrm{occ}}^{3_{+},p}$ & $n_{\mathrm{occ}}^{4_{+},p}$ & $n_{\mathrm{occ}}^{3_{-},p}$ & $n_{\mathrm{occ}}^{4_{-},p}$  \\
			\hline
			$X$ & $29$ & $29$ & $29$ & $29$\\
			$R$ & $28$ & $28$ & $30$ & $30$\\ 
			\hline\hline
		\end{tabular}
		\end{center}
	\end{table}

	We can compute mirror Chern number from the products of rotation eigenvalues listed in Table~\ref{eig_Sr2RuO4}. 
	For the $\chi_{C_4} = -i$ and $\chi_{M_{xy}} = +1$, the mirror chern numbers are 
	\begin{widetext}
		\begin{eqnarray}
		e^{\frac{2\pi i}{4}C^{\sigma = \pm i}_{kx=0 (\pi)}} =(\chi_{C_4})^{2N_{X (R)}^{-\sigma}- N_{\Gamma (Z)}^{-\sigma} - N_{M(A)}^{-\sigma}}\frac{\xi_{\Gamma(Z)}^{+\sigma}\xi_{M(A)}^{+\sigma}}{\zeta_{X(R)}^{+\sigma}}\frac{\xi_{\Gamma(Z)}^{-\sigma}\xi_{M(A)}^{-\sigma}}{\zeta_{X(R)}^{-\sigma}} \\
		(C_{k_z=0}^{+i},C_{k_z=0}^{-i},C_{k_z=\pi}^{+i},C_{k_z=\pi}^{-i})=(2,2,2,2) \mod 4.
		\end{eqnarray}
		
		For the $\chi_{C_4} = +1$ and $\chi_{M_{xy}} = -1$, the mirror chern numbers are 
		\begin{eqnarray}
		e^{\frac{2\pi i}{4}C^{\sigma = \pm i}_{kx=0 (\pi)}} =(\chi_{C_4})^{2N_{X (R)}^{\sigma}- N_{\Gamma (Z)}^{\sigma} - N_{M(A)}^{\sigma}}\left(\frac{\xi_{\Gamma(Z)}^{\sigma}\xi_{M(A)}^{\sigma}}{\zeta_{X(R)}^{\sigma}}\right)^2\\
		(C_{k_z=0}^{+i},C_{k_z=0}^{-i},C_{k_z=\pi}^{+i},C_{k_z=\pi}^{-i})=(2,2,0,0) \mod 4.
		\end{eqnarray}
	\end{widetext}

\begin{table}[H]
	\begin{center}
		\caption{\label{eig_Sr2RuO4}The rotation eigenvalues and the number of occupied bands from DFT.}
		\begin{tabular}{c||c|c}
			\hline\hline
			k-points & $M_{xy} = +i$ sector & $M_{xy} = -i$ sector \\
			\hline
			$\Gamma$ $(0, 0, 0)$& $\xi_{\Gamma}^{+i} = \theta^6$, $N_{\Gamma}^{+i}= 62$ & $\xi_{\Gamma}^{-i} = \theta^{-6}$, $N_{\Gamma}^{-i} = 62$ \\
			$X$ $(\pi, 0, 0)$& $\zeta_{X}^{+i}=-1$, $N_{X}^{+i}= 58$ & $\zeta_{X}^{-i}=-1$, $N_{X}^{-i}= 58$   \\
			$M$ $(\pi, \pi, 0)$& $\xi_{M}^{+i} = \theta^{-7}$, $N_{M}^{+i} = 56$ & $\xi_{M}^{-i} = \theta^7$, $N_{M}^{-i} = 56$  \\
			\hline
			$Z$ $(0,0,\pi)$& $\xi_{Z}^{+i} =1$, $N_{Z}^{+i}= 62$ &$\xi_{Z}^{-i} =1$, $N_{Z}^{-i}= 62$ \\
			$R$ $(\pi, 0, \pi)$& $\zeta_{R}^{+i}=(-i)^2$, $N_{R}^{+i}= 58$ & $\zeta_{R}^{+i}=(-i)^{-2}$, $N_{R}^{-i}= 58$ \\
			$A$ $( \pi, \pi, \pi)$& $\xi_{A}^{+i} =1$, $N_{A}^{+i}= 56$ & $\xi_{A}^{-i} =1$, $N_{A}^{-i}= 56$ \\
			\hline\hline
		\end{tabular}
	\end{center}
\end{table}
	
	\subsection{Character table of point group}
	For reader's convenience, here we reproduce the character tables summarized in Ref.~\onlinecite{PG}. These are the informations necessary to interpret the tables in the previous section.
	
	\begin{table}[H]
		\begin{center}
			\caption{Character table PG $D_{4h}$
				\label{tab:char_D4h}}
			\begin{tabular}{c|c|c|c|c|c|c|c|c|c|c}
				\hline \hline
				Irrep & $E$ & $2C_4$ & $C_2$ &  $2C'_2$  & $2C''_2$ & $\mathcal{I}$ & $2\mathcal{I}C_4$ & $\mathcal{I}C_2$&$2\mathcal{I}C'_2$ &$2\mathcal{I}C''_2$\\
				\hline
				$\Gamma_{6}^{+}$ & $2$ $-2$ & $\sqrt{2}$ $-\sqrt{2}$& $0$ & $0$ & $0$ & $2$ $-2$ & $\sqrt{2}$ $-\sqrt{2}$& $0$ & $0$ & $0$ \\
				$\Gamma_{7}^{+}$& $2$ $-2$ & $-\sqrt{2}$ $\sqrt{2}$& $0$ & $0$ & $0$ & $2$ $-2$ & $-\sqrt{2}$ $\sqrt{2}$& $0$ & $0$ & $0$ \\
				$\Gamma_{6}^{-}$& $2$ $-2$ & $\sqrt{2}$ $-\sqrt{2}$& $0$ & $0$ & $0$ &$-2$ $2$& $-\sqrt{2}$ $\sqrt{2}$& $0$ & $0$ & $0$ \\
				$\Gamma_{7}^{-}$& $2$ $-2$ & $-\sqrt{2}$ $\sqrt{2}$& $0$ & $0$ & $0$ &$-2$ $2$ & $\sqrt{2}$ $-\sqrt{2}$& $0$ & $0$ & $0$ \\
				\hline \hline
			\end{tabular}
		\end{center}
	\end{table}

\begin{table}[H]
	\begin{center}
		\caption{Character table of PG $D_{2h}$
			\label{tab:char_D2h}}
		\begin{tabular}{c|c|c|c|c|c|c|c|c}
			\hline\hline
			Irrep & $E$  & $C_{2}^{z}$ & $C'_{2}$& $C''_{2}$ & $\mathcal{I}$ & $\mathcal{M}_{z}$ & $\mathcal{M}'$ & $\mathcal{M}''$ \\
			\hline
			$\Gamma_{5}^{+}$ & $2$ $-2$ & 0 & 0 & 0 & $2$ $-2$ & 0 & 0 & 0  \\
			$\Gamma_{5}^{-}$ & $2$ $-2$ & 0 & 0 & 0 & $-2$ $2$ & 0 & 0 & 0  \\
			\hline\hline
		\end{tabular}
	\end{center}
\end{table}
	
\begin{table}[H]
	\begin{center}
		\caption{Character table of PG $C_{2h}$
			\label{tab:char_c2h}}
		\begin{tabular}{c|c|c|c|c}
			\hline\hline
			Irrep &  $E$  & $C_2$ & $\mathcal{I}$ & $\mathcal{M}_{z}$\\
			\hline
			$\begin{cases}\Gamma_{3}^{+} \\ \Gamma_{4}^{+} \end{cases} $ &  $\begin{cases} 1 \\1 \end{cases}$ $\begin{cases} -1 \\-1 \end{cases}$& $\begin{cases} i \\-i \end{cases}$ $\begin{cases} -i \\i \end{cases}$ & $\begin{cases} 1 \\1 \end{cases}$ $\begin{cases} -1 \\-1 \end{cases}$& $\begin{cases} i \\-i \end{cases}$ $\begin{cases} -i \\i \end{cases}$ \\
			$\begin{cases}\Gamma_{3}^{-} \\ \Gamma_{4}^{-} \end{cases} $&  $\begin{cases} 1 \\1 \end{cases}$ $\begin{cases} -1 \\-1 \end{cases}$& $\begin{cases} i \\-i \end{cases}$ $\begin{cases} -i \\i \end{cases}$ & $\begin{cases} -1 \\-1 \end{cases}$ $\begin{cases} 1 \\1 \end{cases}$& $\begin{cases} -i \\i \end{cases}$ $\begin{cases} i \\-i \end{cases}$ \\
			\hline \hline
		\end{tabular}
	\end{center}
\end{table}

\begin{table}[H]
	\begin{center}
		\caption{Character table of PG $D_{2d}$
			\label{tab:char_d2d}}
		\begin{tabular}{c|c|c|c|c|c}
			\hline\hline
			Irrep & $E$ & $2\mathcal{I}C_4$ & $C_2$ & $2C'_2$ & $2\sigma_d$\\
			\hline
			$\Gamma_6$ &$2$ $-2$ & $\sqrt{2}$ $-\sqrt{2}$ & $0$ & $0$ & $0$ \\
			$\Gamma_7$ &$2$ $-2$ & $-\sqrt{2}$ $\sqrt{2}$ & $0$ & $0$ & $0$ \\
			\hline \hline
		\end{tabular}
	\end{center}
\end{table}

\begin{table}[H]
	\begin{center}
	\caption{Character table of PG $D_{3d}$
		\label{tab:char_d3d}}
	\begin{tabular}{c|c|c|c|c|c|c}
		\hline\hline
		Irrep & $E$ & $2C_3$ & $3C'_2$ & $I$ & $2S_6$ & $3\sigma_d$\\
		\hline
		$\Gamma_{4}^{+}$ &$2$ $-2$ & $1$ $-1$ & $0$ &$2$ $-2$ & $1$ $-1$ & $0$\\
		$\Gamma_{5}^{+}$ &$1$ $-1$ & $-1$ $1$ & $i$ $-i$ &$1$ $-1$ & $-1$ $1$ & $i$ $-i$ \\
		$\Gamma_{6}^{+}$ &$1$ $-1$ & $-1$ $1$ & $-i$ $i$ &$1$ $-1$ & $-1$ $1$ & $-i$ $i$ \\
		$\Gamma_{4}^{-}$ &$2$ $-2$ & $1$ $-1$ & $0$ &$-2$ $2$ & $-1$ $1$ & $0$\\
		$\Gamma_{5}^{-}$ &$1$ $-1$ & $-1$ $1$ & $i$ $-i$ &$-1$ $1$ & $1$ $-1$ & $-i$ $i$ \\
		$\Gamma_{6}^{-}$ &$1$ $-1$ & $-1$ $1$ & $-i$ $i$ &$-1$ $1$ & $1$ $-1$ & $i$ $-i$ \\
		\hline \hline
	\end{tabular}
	\end{center}
\end{table}

\begin{table}[H]
	\begin{center}
		\caption{Character table of PG $D_{3}$
			\label{tab:char_d3}}
		\begin{tabular}{c|c|c|c}
			\hline\hline
			Irrep & $E$ & $2C_3$ & $3C'_2$ \\
			\hline
			$\Gamma_{4}$ &$2$ $-2$ & $1$ $-1$ & $0$ \\
			$\Gamma_{5}$ &$1$ $-1$ & $-1$ $1$ & $i$ $-i$  \\
			$\Gamma_{6}$ &$1$ $-1$ & $-1$ $1$ & $-i$ $i$ \\
			\hline \hline
		\end{tabular}
	\end{center}
\end{table}

\begin{table}[H]
	\begin{center}
		\caption{Character table of PG $C_{4v}$
			\label{tab:char_c4v}}
		\begin{tabular}{c|c|c|c|c|c}
			\hline\hline
			Irrep & $E$ & $2C_4$ & $C_2$ & $2\sigma_v$ & $2\sigma_d$\\
			\hline
			$\Gamma_6$ &$2$ $-2$ & $\sqrt{2}$ $-\sqrt{2}$ & $0$ & $0$ & $0$ \\
			$\Gamma_7$ &$2$ $-2$ & $-\sqrt{2}$ $\sqrt{2}$ & $0$ & $0$ & $0$ \\
			\hline \hline
		\end{tabular}
	\end{center}
\end{table}

\begin{table}
	\begin{center}
		\caption{Character table of PG $C_{2v}$
			\label{tab:char_c2v}}
		\begin{tabular}{c|c|c|c|c}
			\hline\hline
			Irrep & $E$ & $C_2$ & $\sigma_v$ & $\sigma'_{v}$\\
			\hline
			$\Gamma_{5}$ &$2$ $-2$ & $0$ & $0$ & $0$\\
			\hline \hline
		\end{tabular}
	\end{center}
\end{table}

\bibliography{material_search}

\end{document}